\documentclass[entropy,article,accept,moreauthors,pdftex]{Definitions/mdpi} 

\firstpage{1} 
\pubvolume{xx}
\issuenum{1}
\articlenumber{5}
\pubyear{2019}
\copyrightyear{2019}
\history{Received: date; Accepted: date; Published: date}
\updates{yes} 

\usepackage[utf8]{inputenc}
\usepackage{graphicx}
\usepackage{caption}
 \usepackage[labelformat=simple]{subcaption} 

\DeclareCaptionLabelFormat{subcaptionlabel}{\normalfont(\textbf{#2}\normalfont)}
\usepackage{color}

\usepackage{amsmath}
\usepackage{amssymb}
\usepackage{wrapfig}
\usepackage{comment}
\usepackage{gensymb}
\usepackage{float}
\usepackage{tabularx}
\usepackage{algorithm}
\usepackage{algorithmic}
\usepackage{amssymb,amsmath}
\usepackage{relsize}
\usepackage{setspace}
\usepackage[english]{babel}
\usepackage{array}
\usepackage{multirow}
\usepackage{tabularx,booktabs} 
\usepackage[export]{adjustbox}
\usepackage{url}
\usepackage{array}
\usepackage{array, booktabs}
\setitemize{parsep=6pt,itemsep=0pt,leftmargin=*,labelsep=5.5mm}
\setenumerate{parsep=6pt,itemsep=0pt,leftmargin=*,labelsep=5.5mm}
\setlist[description]{itemsep=0mm}

\DeclareUnicodeCharacter{2010}{-}

\newcommand{\bmark}[1]{#1}

\Title{Multivariate Pointwise Information-Driven Data Sampling and Visualization}

\Author{ {Soumya Dutta} 
 *\orcidA{}, Ayan Biswas, and James Ahrens}

\AuthorNames{Soumya Dutta, Ayan Biswas, and James Ahrens}

\address[1]{%
  {Los Alamos National Laboratory, Los Alamos, NM - 87545, USA}
 }

\corres{\hangafter=1 \hangindent=1.05em \hspace{-0.82em} Correspondence: sdutta@lanl.gov}

\abstract{
\textls[-20]{With increasing computing capabilities of modern supercomputers, the size of the data generated from the scientific simulations is growing rapidly. As a result}, application~scientists need effective data summarization techniques that can reduce large-scale multivariate spatiotemporal data sets while preserving the important data properties so that the reduced data can answer domain-specific queries involving multiple variables with sufficient accuracy. While~analyzing complex scientific events, domain experts often analyze and visualize two or more variables together to obtain a better understanding of the characteristics of the data features. Therefore,~data~summarization techniques are required to analyze multi-variable relationships in detail and then perform data reduction such that the important features involving multiple variables are preserved in the reduced data. To achieve this, in this work, we propose a data sub-sampling algorithm for performing statistical data summarization that leverages pointwise information theoretic measures to quantify the statistical association of data points considering multiple variables and generates a sub-sampled data that preserves the statistical association among multi-variables. Using such reduced sampled data, we show that multivariate feature query and analysis can be done effectively. The efficacy of the proposed multivariate association driven sampling algorithm is presented by applying it on several scientific data sets.
}


\begin{document}

\section{Introduction}

The size of the scientific data sets is increasing rapidly with ever-increasing computing capabilities. Modern-day supercomputers can generate data in the order of petabytes and soon we will enter the era of exascale computing~\cite{doe_exascale_report, JSFI78}. As~the size of the data sets keeps growing, traditional analysis and visualization techniques using full resolution raw data will soon become prohibitive since storing, parsing, and~analyzing the full resolution raw data will not be a viable option anymore~\cite{Jim_SC_cinema, adr, explorable_images, dutta_vis16}. This is primarily due to the gap between the disk I/O speed and the data generation speed. Recent studies show that the velocity at which data is being generated is significantly higher compared to the speed at which the data can be stored into the disks~\cite{Jim_SC_cinema,dutta_vis16,woodring_mpas_insitu}. Therefore, only a small subset of the data can be moved to the permanent storage for exploratory post-hoc~analysis. 

Since the data triage is going to be a necessary step in the near future to facilitate flexible, exploratory, and~timely post-hoc analysis, the~important question now is that how the data should be reduced such that the stored subset of the data retains important information about the scientific simulations and is able to answer the queries performed by domain scientists. Large-scale simulations nowadays output data sets containing multiple variables over time. During~the analysis phase of such multivariate data sets, application scientists often explore two or more variables simultaneously instead of looking at each variable individually to investigate the modeled physical phenomena in their simulation. For~example, to~understand the complex turbulent chemical processes of combustion, scientists have acknowledged that the exploration of multiple variables together is essential and analyzing variables individually is not sufficient~\cite{akiba_combustion}. By~exploring multiple variables together, scientists~found that during the combustion process, high valued hydroxyl regions are located where the stoichiometric mixture fraction isosurfaces are convoluted. Similarly, in~the climate science domain, while studying the impact of a hurricane in a region, climatologists often look at variables like Pressure and Velocity together since the hurricane eye, an~important feature of the hurricane system, can be located by querying for the regions that have low Pressure and low to moderate Velocity~\cite{gosink_1, Hazarika_copula_multivar}. Therefore, it is essential to preserve the relationship among variables while performing data reduction such that multivariate feature analysis, as~mentioned above, can be performed during the post-hoc analysis~phase.

As previous researchers have shown, relationships among multiple variables in the scientific data sets can be complicated~\cite{xiaotong_association, vis13_biswas, Janicke2007} and so, effective summarization of such large multivariate data sets is a challenging task. Since not all the data values are equally important, the~application scientists often employ query-driven analysis techniques where a different range of values of multiple variables are queried and analyzed simultaneously~\cite{gosink_1, Hazarika_copula_multivar,query_driven}. The~query can involve two or more variables and the result of the query on multiple variables can be combined using boolean clauses such as AND/OR.  The~query ranges can also be further refined as the experts gain knowledge from the exploratory analyses. As~a result, a~primary requirement of such summarization process will be to preserve the salient regions (i.e., regions where multiple variables demonstrate specific trends) of the data with high fidelity such that query-driven multivariate analysis and visualization can be done efficiently using the reduced subset of the data. Scientists in the past have proposed several data reduction approaches such as distribution-based~\cite{dutta_vis16, Hazarika_copula_multivar,ko_chih_2,ko_chih_1,dutta_slic_pvis}, image-based~\cite{Jim_SC_cinema, explorable_images}, wavelet compression based~\cite{clyne2007interactive, 8048933, 7348075}, data compression by functional approximations~\cite{ISABELA} etc. However, a~majority of these techniques perform data reduction on each variable individually and overlook the relationship among variables. Therefore, even though the above techniques have achieved promising results for summarizing data variables individually, multivariate relationship-aware data summarization techniques that can facilitate query-driven visual analysis during the post-hoc analysis phase are still~missing.      

In this work, we propose a multivariate data summarization technique that selects a small fraction of the original data set (i.e., subsamples the data) considering multiple variables simultaneously and picks data points with higher fidelity from the regions where the selected variables show a strong association. The~benefits of sampling-based data representations for a single variable has been shown by the researchers in the past~\cite{isav_2018_sampling, tzu_pvis, Woodring_sampling, yu_Su_bitmap_sampling}. In~this work, we pursue a novel sampling-based data summarization scheme and extend the sampling capability to the multi-variable domain. The~proposed sampling technique uses information theoretic measure pointwise mutual information (PMI) \cite{Church_PMI_Main_CL89} (for bivariate sampling) and its multivariate generalization specific correlation~\cite{VandeCruys_MultiVarPMI_DiSCo11} to assign importance to each data point based on their multivariate associativity. Given a data point containing multiple values from different variables, pointwise mutual information allows us to quantify the strength of associativity of the data point. We leverage statistical distribution-based data sampling technique to produce a sub-sampled data set where the data points are sampled according to the strength of their statistical association values. Using pointwise statistical association as a multivariate sampling criterion, we generate a sub-sampled data where more samples are taken from regions with strong associativity and fewer samples are taken from regions that demonstrate weak dependence among the selected variables. \bmark{We demonstrate the usefulness of the proposed sampling algorithm by applying it to several scientific data sets. Furthermore, to~study the effectiveness of the proposed method, we~compare it both qualitatively and quantitatively to the traditional random sampling technique. In~our study, we~demonstrate the applications where the proposed method is desirable over random sampling and also report its potential limitations compared to the random sampling.}  Therefore,~our~contributions to this work are threefold:

\begin{enumerate}[leftmargin=*,labelsep=5mm]
\item We propose a new multivariate association-driven data sampling algorithm for large-scale data~summarization.
\item Given a user-specified sampling fraction, we use pointwise information measures and statistical distribution-based sampling techniques to generate a sub-sampled data that preserves the important multivariate features. 
\item We perform a detailed qualitative and quantitative study to demonstrate the efficacy of the proposed sampling scheme.
\end{enumerate}

\section{Related~Works}
\vspace{-6pt}

\subsection{Information Theory in~Visualization}

In computer graphics and visualization, information theory~\cite{Cover2006,Shannon:2001,shannonTheory} has been used to solve a variety of problems in the past decades. In~particular, the~concepts of entropy and mutual information (MI) have been very popular for applications that use ideas ranging from information quantification to information overlap. Viola~et~al.~\cite{Viola2006} used MI for finding the best viewpoints for a given scene and achieving its smooth transition. MI has also been applied successfully for data fusion and image registration~\cite{MI_registration_1995,hill2001,WellsIII199635,multimodal_reg_MI,Pluim03mibasedregistration}. Previously, MI has further been used by researchers for understanding scene~\cite{Feixas99aninformation} and shape complexity~\cite{1563243}, as~well as understanding and analyzing mesh properties~\cite{Feixas06aunified}. For~scalar datasets, Bruckner~et~al.~\cite{bruckner_isosurface_similarity_maps} used MI to identify the isosurface similarities. Wei~et~al.~\cite{Wei:2013} used MI to compute representativeness of isosurfaces via a level-sets for volumetric data. Bramon~et~al.~\cite{Bramon:ObservationChannel:Evis:2013} used MI between data values and color pixels to compute information transfer. MI has also been used on the vector datasets for the selection of streamlines~\cite{ma_importance, Tao_2013}. In~recent years, decompositions of MI such as specific mutual information (SMI) and pointwise mutual information (PMI) have become popular for characterizing properties of individual scalars of a dataset. For~fusing data from multiple sources, Bramon~et~al.~\cite{multimodal_data_fusion} used SMI and showed the trade-off of different fusion operations. Biswas~et~al.~\cite{vis13_biswas} used entropy-based methods for salient variable selection and applied SMI for identification of salient isosurfaces from such variables. SMI has also been used for transfer function design~\cite{6516552}. PMI was used by Haidacher~et~al.~\cite{haidacher-2008-vcbm} for understanding and extracting inverse associations from the datasets. PMI was also used by Dutta~et~al.~\cite{Dutta_pmi} for exploration of time-varying multivariate datasets.
For more detailed reviews regarding use of information theory in visualization, we refer the readers to~\cite{info_theory_book, chen_janicke_infotheory,4459862,sbert09informationTheory,chaoli_infotheory_entropy}. 

\subsection{Sampling for Data Analysis and~Visualization}

Sampling has been a very widely popular approach for selecting a subset of values from a given population.  Visualization researchers have used sampling primarily in the context of data reduction such that fast data analysis and visualization can be performed. A~stratified random sampling based scheme was proposed by Woodring~et~al.~\cite{Woodring_sampling} where the authors used cosmology simulation as their application and enabled interactive visualization of the large-scale data. Using bitmap indices and information entropy, Wei~et~al.~\cite{tzu_pvis} extended the standard stratified random sampling for in~situ data reduction. Previous to this work, Su~et~al.~\cite{yu_Su_bitmap_sampling} used bit map indices for sampling to enable fast user queries on the datasets. Recently, visualization aware sampling methods are starting to gain more importance~\cite{viz_aware_sampling}. In~this work, Park~et~al.~\cite{viz_aware_sampling} generated samples for scatter plot and map plot to still retain important visualization properties. Nguyen and Song~\cite{centrality_cluster_based_sampling} proposed a sampling approach that used the centrality-driven clustering for getting higher performance and quality over existing simple random sampling methods. Use of information theory has also been an important direction for researchers~\cite{entropy_max_1,entropy_max_2, entropy_max_3} while finding the optimal subset of the original data. Biswas~et~al.~\cite{isav_2018_sampling} used similar ideas of entropy maximization for inventing an in~situ data-driven sampling scheme that can preserve important properties of the data. They used probability distribution of the original datasets to find important samples in a generic sampling~scheme.

\subsection{Multivariate Data Analysis and~Visualization}

Multivariate data analysis and visualization have historically been very important for researchers. A~study of local correlation coefficients among multiple variables for visualization purposes was provided by Sauber~et~al.~\cite{Sauber2006}. For~multivariate datasets, Bethel~et~al.~\cite{gosink_query} further used correlations for enabling query-driven analysis and visualization. Such query-driven methods were extended by Gosink~et~al.~\cite{gosink_1} where they used local statistical data properties. J\"{a}nicke~et~al.~\cite{Janicke2007} used local statistical complexities for finding out the interesting regions of multivariate datasets. Despite the works, multivariate data analysis remains a challenging task and there exists no sampling approach that specifically uses multivariate information for data reduction purposes. For~a detailed overview of this topic, we refer the interested readers to the works of Wong~et~al.~\cite{Wong1994} and Fuchs~et~al.~\cite{Fuchs}.

\section{Method}

In this section, we briefly discuss existing generic sampling techniques and then introduce the proposed multivariate association-driven sampling algorithm. Statistical data sampling has been shown to be effective in summarizing and analyzing data sets by researchers in the past~\cite{isav_2018_sampling, sampling_book}. One key advantage of sampling-based data summarization over sophisticated data modeling-based approaches is that the sampling techniques keep a true subset of representative points selected from the original raw data. Therefore, the~observed variable values at these selected point locations are accurate, whereas using statistical models such as distributions, the~values at such spatial locations will have to be inferred from the model and may contain uncertainties. Furthermore, the~sub-sampled data sets can be analyzed and visualized directly without any further processing, whereas, data summarized by statistical models often need to perform reconstruction of data first which may be an expensive operation. Also, sampling techniques are expected to be computationally less expensive compared to the sophisticated statistical modeling algorithms where iterative processes are employed for model parameter estimation. To~generate a representative sub-sampled data, a~simple technique is to sample the data at regular intervals. This technique is called regular sampling. Regular sampling does not consider any data properties while selecting samples and due to the regular nature of sample selection, it produces artifacts and discontinuities during sample-based visual analysis~\cite{isav_2018_sampling, dutta_slic_pvis}.

\subsection{Random~Sampling} \label{random_sampling}

Another well-known and widely used sampling technique is random sampling.  Random sampling for data summarization has been a promising approach among scientists and is being used extensively since the sub-sampled data set preserves the overall distribution of the data along with statistical quantities such as mean, standard deviation, etc. In~the random sampling algorithm, each data sample has an equal probability for getting picked and hence it does not consider any variable relationship while selecting points. Random sampling operates independently on each data point and can be performed by applying the rejection sampling algorithm~\cite{albert2009bayesian} for selecting data points uniformly randomly. Given a user-specified sampling fraction $\alpha$ ($0 < \alpha < 1$), for~each data point we generate a sample $s$ ($0 \leq s \leq 1$)  drawn from a standard uniform distribution $U(0, 1)$. If~$s < \alpha$ then the current data point is selected. Since the random sampling technique does not consider variable relationships while selecting data points, the~sampled output does not warrant any feature preservation. In~the following section, we provide a new multivariate sampling scheme that exploits statistical association among multiple variables while sampling data points and hence is able to preserve important features in the data with higher~accuracy.

\subsection{Proposed Multivariate Statistical Association-Driven~Sampling} \label{pmi_sampling}
Our primary goal in this work is to develop a data sub-sampling algorithm that considers multiple variables such that the reduced sampled data set preserves the statistical association among those variables with higher fidelity. To~achieve this, the~proposed sampling scheme samples densely from regions where a subset of selected data variables show a strong statistical association. It has been shown previously~\cite{xiaotong_association, Dutta_pmi} that such associative regions often represent a multivariate feature in scientific data sets where a range of values of multiple variables tends to co-occur frequently. Therefore, considering~two variables and a pair of the scalar value selected from them, the~existence of a strong statistical association between the value pair can be comprehended if they demonstrate high co-occurrence and the distribution of these value pairs in the spatial domain represents a joint statistical feature. The~proposed sampling algorithm samples data points by following such multivariate statistical association and selects more samples from such feature regions such that the feature regions are well preserved. At~the end of the sampling process, the~output is an unstructured non-uniform set of points. Before~we present the details of the sampling technique, first we introduce the information theoretic measure pointwise mutual information that allows us to quantify multivariate associativity for each data point and then present the sampling algorithm that uses pointwise information as the sample selection~criterion.

\subsubsection{Multivariate Pointwise Information~Characterization} \label{pmi_subsection}
To perform multivariate statistical association driven sampling, characterization of importance of each data point is essential. In~the multivariate data set, each data point has a value tuple consisting of values from each data variable associated with it.  Consider a bivariate example where $X$ and $Y$ are two variables and at each data point, we have a value pair $(x,y)$. Here, $x$ is a specific value of variable $X$ and $y$ for variable $Y$.  For~each such value pair, the~shared information needs to be quantified so that we can identify data points (associated with each value pair) having a higher statistical association. Information theoretic measure pointwise mutual information (PMI) allows us to quantify such shared information. Given two random variables $X$ and $Y$, if~$x$ is an observation of $X$ and $y$ for $Y$, then the PMI value for the value pair $(x,y)$ is expressed as:
\begin{equation} \label{eq:pointwise_mutual_info}
 PMI(x,y) = \log{\frac{p(x,y)}{p(x)p(y)}} 
\end{equation} 
where $p(x)$ is the probability of a particular occurrence $x$ of $X$, $p(y)$ is the probability of $y$ of variable $Y$ and, $p(x,y)$ is their joint probability. PMI was first introduced by Church and Hanks~\cite{Church_PMI_Main_CL89} for the quantification of word association directly from computer readable corpora. When $p(x,y) > p(x)p(y)$, $PMI(x,y) > 0$, which means $x$ and $y$ have higher information sharing between them. If~$p(x,y) < p(x)p(y)$, then $PMI(x,y) < 0$ indicating the two observations follow complementary distribution. When $x$ and $y$ do not have any significant information overlap then $p(x,y) \approx p(x)p(y)$ and $PMI(x,y) \approx$ 0. In~this case, $x$ and $y$ are considered as statistically independent. Note that, the~mutual information $I(X;Y)$ yields the expected PMI value over all possible instances of variable $X$ and $Y$ \cite{VandeCruys_MultiVarPMI_DiSCo11}.
\begin{equation}
 I(X;Y) =  E_{(X,Y)}[PMI(x,y)] \label{eq:expected_pointwise_mutual_info}
\end{equation} 

By estimating PMI values for each value pair, the~shared information content for each value pair can be quantified and since each data point is associated with a value pair, we can now quantify the importance of each data point by analyzing its PMI value. If~the value pair associated with a data point has high PMI value, then it is considered more important in our sampling algorithm since high PMI value indicates strong co-occurrence and therefore strong statistical association~\cite{Church_PMI_Main_CL89}. Furthermore, previous researchers also showed that such high PMI valued regions in multivariate data sets generally correspond to multivariate features~\cite{Dutta_pmi} where multiple variables demonstrate a strong statistical association. Hence, using PMI the strength of statistical association for each data point can be quantified~effectively.

\begin{figure}[H]
\centering
\begin{subfigure}[H]{0.28\linewidth}
     \centering
    \includegraphics[width=\linewidth]{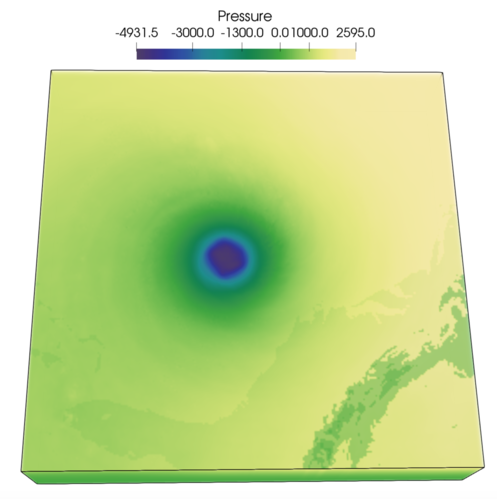}
    \caption{Pressure~field~visualization.}
    \label{fig:Isabel_P}
\end{subfigure}
~
\begin{subfigure}[H]{0.28\linewidth}
     \centering
    \includegraphics[width=\linewidth]{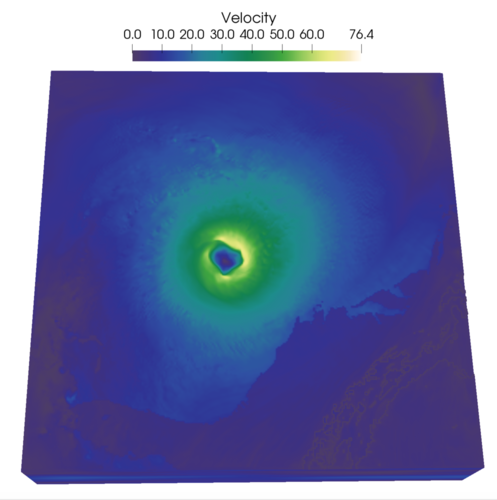}
    \caption{Velocity field~visualization.}
    \label{fig:Isabel_VEL}
\end{subfigure}
\caption{ {Visualization} 
 of Pressure and Velocity field of Hurricane Isabel data set. The~hurricane eye at the center of Pressure field and the high velocity region around the hurricane eye can be~observed.}
\label{fig:Isabel_example}
\end{figure}

\begin{figure}[H]
\centering
\begin{subfigure}[H]{0.25\linewidth}
     \centering
    \includegraphics[width=\linewidth]{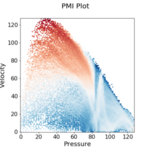}
    \caption{PMI plot of Pressure and Velocity~field.}
    \label{fig:Isabel_pmi_plot_P_VEL}
\end{subfigure}
~
\begin{subfigure}[H]{0.25\linewidth}
     \centering
    \includegraphics[width=\linewidth]{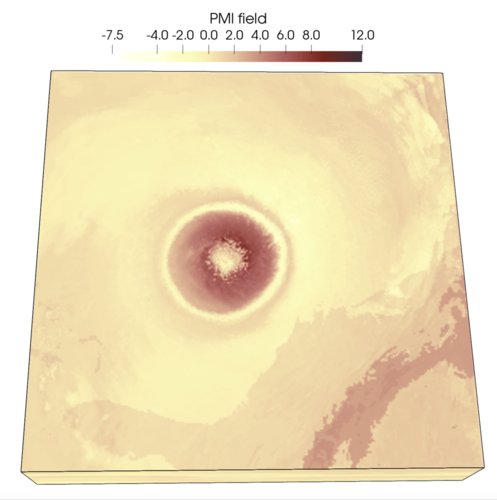}
    \caption{PMI volume of Pressure and Velocity~field.}
    \label{fig:PMI_field_P_VEL}
\end{subfigure}
\caption{ {PMI} computed from Pressure and Velocity field of Hurricane Isabel data set is visualized. Figure~\ref{fig:PMI_example}a shows the 2D plot of PMI values for all value pairs of Pressure and Velocity, Figure~\ref{fig:PMI_example}b provides the PMI field for analyzing the PMI values in the spatial domain. It can be seen that around the hurricane eye, the~eyewall is highlighted as high PMI-valued region which indicates a joint feature in the data set involving Pressure and Velocity~field.}
\label{fig:PMI_example}
\end{figure}

In Figure~\ref{fig:Isabel_example}, renderings of two scalar fields, Pressure, and~Velocity, of~Hurricane Isabel data set is shown to illustrate the usefulness of PMI values in characterizing multivariate association. As~PMI values can be computed for each value pair of two variables, a~2-D PMI plot can be obtained where the PMI values of all value pairs can be studied. Such a PMI plot for Pressure and Velocity variables is shown in Figure~\ref{fig:PMI_example}a. The~x-axis of the plot contains values of Pressure and the y-axis contains values of Velocity (Note that, the~computation of PMI values are done on a histogram space and hence, the~axes in plot \ref{fig:PMI_example}a show the bin ids. The~scalar value that represents the bin center for each bin can be trivially computed from the range of data values for each variable). The~white regions in the plot represent value pairs with zero PMI value. As~we can see from this plot, the~lower values of Pressure and moderate to high values of Velocity has higher PMI values (seen from the red regions in the plot). \mbox{Now, since each} spatial point in the data set has a Pressure and Velocity value associated with it, we can easily construct a new scalar field, called PMI field as suggested in~\cite{Dutta_pmi} by assigning the PMI value at each grid point. Visualization of such a PMI field is provided in Figure~\ref{fig:PMI_example}b. This PMI field can be considered as an association field between Pressure and Velocity, and~high PMI valued regions in this field will indicate regions that demonstrate a higher statistical association. From~Figure~\ref{fig:PMI_example}b, we can observe that the high PMI valued data points are located on the eyewall of the hurricane (the dark reddish region at the center in Figure~\ref{fig:PMI_example}b). This is an important feature in Isabel data set and the eyewall of a hurricane typically represents the region where the most destructive high-velocity wind exists. Hence, the~above discussion demonstrates the effectiveness of PMI in quantifying statistical association for each data point in the spatial domain. To~preserve such statistical associative features while performing data sampling, in~this work, we use PMI values as the sampling criterion for determining whether data point will be selected or not. The~proposed sampling algorithm selects data points densely where the data points have high PMI values. In~Section~\ref{sampling_subsection} the proposed pointwise information-driven sampling algorithm is presented in~detail.

\unskip

\unskip

\subsubsection{Generalized Pointwise~Information} \label{general_pmi}
In the above section, we introduced the information theoretic measure PMI which allows quantification of statistical association for each data point which is applicable for two variables only. In~this section, we provide a generalized extension of PMI which enables us to use more than two variables while analyzing statistical association for data points for sampling. Watanabe in their work~\cite{Watanabe:Total_Correlation:1960} first quantified the total shared information among multiple variables as:
\begin{equation}
 I(X_1,X_2,..,X_n) =  \sum_{x_1 \in X_1, x_2 \in X_2, ..., x_n \in X_n}^{} p(x_1,x_2,...,x_n)log\frac{p(x_1,x_2,...,x_n)}{p(x_1)p(x_2)...p(x_n)} \label{eq:total_correlation}
\end{equation} 

The quantity $I(X_1,X_2,..,X_n)$ is termed as \textit{total correlation} where $n$ represents the number of variables, $p(x_i)$ represents the probability of a specific value $x_i$ of $i^{th}$ variable $X_i$, and~$p(x_1,x_2,...,x_n)$ indicates the joint probability of the value tuple $(x_1,x_2,...,x_n)$. Note that, total correlation quantifies the total shared information among multiple variables and does quantify importance for each data point. Later,~from~the above definition, Tim Van de Cruys defined a new information theoretic measure called \textit{specific correlation} \cite{VandeCruys_MultiVarPMI_DiSCo11}, which is analogous to PMI for multiple variables:
\begin{equation}
SI(x_1,x_2,..,x_n) = log\frac{p(x_1,x_2,...,x_n)}{p(x_1)p(x_2)...p(x_n)} \label{eq:pmi_multivar}
\end{equation} 

As can be seen, specific correlation ($SI(x_1,x_2,...,x_n)$) presented in Equation~(\ref{eq:pmi_multivar}) is a natural extension of bi-variate PMI depicted in Equation~(\ref{eq:expected_pointwise_mutual_info}). Specific correlation measure follows similar statistical properties as PMI discussed above, and~higher values of specific correlation indicate stronger statistical~association.

\subsubsection{Pointwise Information-Guided Multivariate~Sampling} \label{sampling_subsection}
In the above section, we presented pointwise mutual information (PMI) and a generalized extension of it which allows us to quantify the importance of each data point in terms of their statistical association considering multiple variables. In~the following, we propose a novel sampling algorithm, which uses such pointwise information measure as a sampling criterion such that multivariate statistical association driven sampling can be done. Using pointwise information assigned to each data point, the~proposed sampling scheme accepts data points with higher likelihood when the pointwise information value is high indicating a high statistical association. As~a result, the~sub-sampled data set is able to preserve the multivariate association with higher fidelity and facilitates efficient visual query and~analysis. 

Given a user-specified sampling fraction $\alpha$ ($0 < \alpha < 1$) as an input parameter to the sampling algorithm, the~proposed sampling scheme outputs a sub-sampled data set containing $n = \alpha \times N (n < N)$ points where $N$ is the total number of data points. In~order to sample data points using their pointwise information, in~this work, we use a multivariate distribution-based approach. Consider a bivariate example, where $X$ and $Y$ are two data variables using which sampling will be done. Firstly,~the~joint probability distribution of these two variables is estimated using a histogram. The~univariate histograms of variable $X$ and $Y$ can be estimated by marginalizing the joint histogram. Since we are considering two variables to describe the sampling algorithm in this example, the~joint histogram, in~this case, is a 2D histogram. Each bin center in this 2D histogram represents a value pair $(x_i,y_j)$ for the two variables for the bin $(i,j)$. Also, given the joint and univariate distributions of variables $X$ and $Y$, the~PMI value for each value pair $(x_i,y_j)$ for the bin $(i,j)$ ($0 \leq i \leq B-1~and~0 \leq j \leq B-1$ where B is the number of bins) can be estimated by following Equation~(\ref{eq:pointwise_mutual_info}). Therefore,~a~PMI value is assigned to each 2D histogram bin. Since each bin contains multiple data points, all the data points belonging to a specific 2D histogram bin is assigned with the PMI value of the current bin. In~this way, all the data points are assigned a PMI value, same as its bin center's PMI value. Finally, the~normalized PMI values assigned to each 2D histogram bin is treated as the acceptance probability for the data points in that bin. For~example, if~a bin in the 2D histogram has normalized PMI value $0.7$, then $70\%$ of the data points from this bin will be selected. Since the sampling criterion, i.e.,~the normalized PMI value represents the strength of statistical association for multivariate data, this sampling technique will accept more sample points from the 2D histogram bins where the value of PMI is high. As~a result, the~final sub-sampled data set will preserve the strong statistically associative regions with higher~fidelity.

The above sampling technique is similar in spirit to the importance sampling algorithms in statistics~\cite{Doucet2000, Earl_importance_sampling} where Monte Carlo methods are used for estimating statistical expectations of one distribution by sampling from another distribution. In~our algorithm, we use PMI values as the importance criterion while selecting data points. The~selection of points from each 2D histogram bin is done using rejection sampling method~\cite{albert2009bayesian}. To~determine whether to select a data point belonging to bin $(i,j)$ having normalized PMI value $pmi(i,j)$, first we generate a sample $s$ drawn from a standard uniform distribution $U(0, 1)$. If~$s < pmi(i,j)$ then the data point is selected. This sample selection process is repeated for all the data points for each bin. It is to be noted that, since the sampling algorithm is expected to store $n$ data points according to the user-specified sampling fraction $\alpha$, the~proposed sampling scheme first scales the normalized PMI values by applying a scaling factor $\gamma$ such that the following condition is satisfied:
\begin{equation}
 \sum_{i=0,j=0}^{B-1,B-1} \gamma*pmi(i,j)*f(i,j) = n  \label{eq:sampling_condition}
\end{equation}
where $f(i,j)$ is the frequency of the 2D histogram bin $(i,j)$ and $B$ is the number of histogram bins. \bmark{The scaling factor $\gamma$ scales the PMI values to ensure that the total number of data points selected finally meets the desired sampling fraction $\alpha$. Estimation of $\gamma$ for a given sampling fraction $\alpha$ is straightforward. First we compute the number of data points $n^\prime$ that would be picked without any scaling by evaluating $ n^\prime = \sum_{i=0,j=0}^{B-1,B-1} pmi(i,j)*f(i,j)$. If~$n^\prime < n$ then scaling factor $\gamma = n/n^\prime$, else scaling factor $\gamma = n^\prime/n$. Please note that the above sampling algorithm naturally extends to multi-variable domain (i.e., more than two variables) and for the characterization of the statistical association in such cases, the~generalized definition of pointwise statistical association, i.e.,~the specific correlation measure presented in Equation~(\ref{eq:pmi_multivar}) can be used without the loss of any generality.}

\begin{figure}[H]
\centering
\begin{subfigure}[t]{0.27\linewidth}
     \centering
    \includegraphics[width=\linewidth]{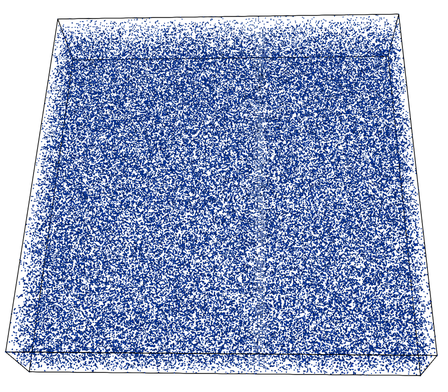}
    \caption{Randomly sampled~points.}
    \label{fig:isabel_random_sampled_3_percent}
\end{subfigure}
~
\begin{subfigure}[t]{0.27\linewidth}
     \centering
    \includegraphics[width=\linewidth]{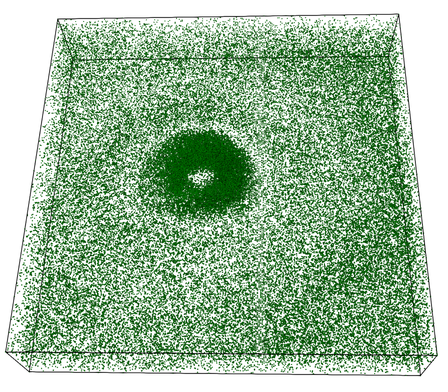}
    \caption{Proposed statistical association driven sampled~points.}
    \label{fig:isabel_pmi_sampled_3_percent}
\end{subfigure}
\caption{Sampling result on Isabel data set when Pressure and Velocity variables are used. Figure~\ref{fig:Isabel_sampling_example}a shows results of random sampling and Figure~\ref{fig:Isabel_sampling_example}b shows results of the proposed pointwise information driven sampling results for sampling fraction $0.03$. By~observing the PMI field presented in Figure~\ref{fig:PMI_example}b, it can be seen that the proposed sampling method samples densely from the regions where statistical association between Pressure and Velocity is stronger (Figure \ref{fig:Isabel_sampling_example}b).}
\label{fig:Isabel_sampling_example}
\end{figure}

\begin{figure}[t]
\centering
\begin{subfigure}[t]{0.2\linewidth}
     \centering
    {\includegraphics[width=\linewidth]{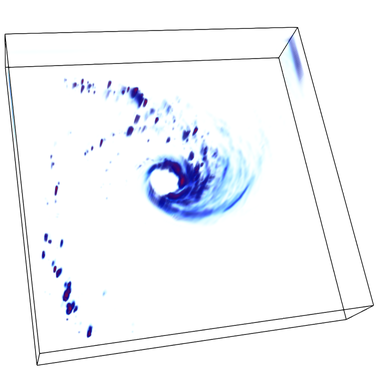}}
    \caption{Visualization of QGraup~field.}
    \label{fig:qgraup_raw}
\end{subfigure}
~~
\begin{subfigure}[t]{0.2\linewidth}
     \centering
    {\includegraphics[width=\linewidth]{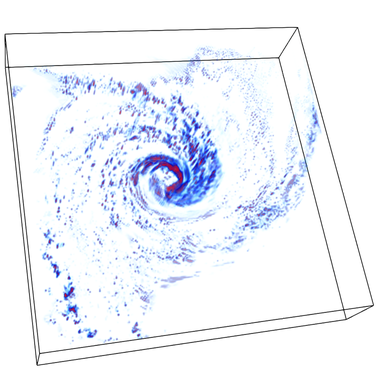}}
    \caption{Visualization of QCloud~field.}
    \label{fig:qcloud_raw}
\end{subfigure}
~~
\begin{subfigure}[t]{0.2\linewidth}
     \centering
    {\includegraphics[width=\linewidth]{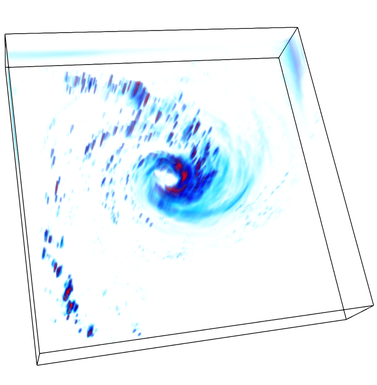}}
    \caption{Visualization of Precipitation~field.}
    \label{fig:precipitation_raw}
\end{subfigure}
~
\begin{subfigure}[t]{0.2\linewidth}
     \centering
    {\includegraphics[width=\linewidth]{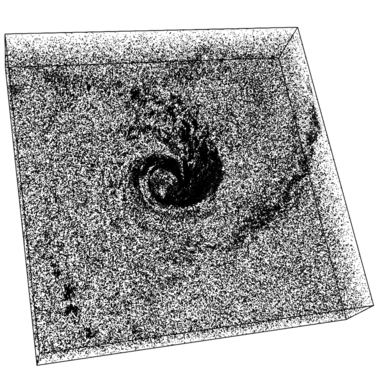}}
    \caption{Sample points selected by the proposed multivariate sampling~algorithm.}
    \label{fig:isabel_pmi_sampled_qgraup_qcloud_precip_5}
\end{subfigure}
\caption{Sampling result for Isabel data set when three variables (QGraup, QCloud, and~Precipitation) are used to perform sampling. In~this case, the~generalized specific correlation measure presented in Equation is used to compute multivariate associativity for the data points considering all three variables. Figure~\ref{fig:isabel_three_var_example}a--c show the rendering of QGraup, QCloud, and~Precipitation fields respectively. Figure~\ref{fig:isabel_three_var_example}d presents the rendering of sampled data points when the proposed multivariate sampling algorithm is applied to these three variables. It can be seen that the cloud and the rain bands show stronger statistical association among three variables and hence are sampled densely. The~sampling fraction used in this example is $0.05$.}
\label{fig:isabel_three_var_example}
\end{figure}

An example of the proposed sampling algorithm is provided in Figure~\ref{fig:Isabel_sampling_example} using Isabel data set where Pressure and Velocity variables are used while sampling data points. The~sampling fraction for this example is set to $\alpha=0.03$, i.e.,~$3\%$ of total data points are selected in the sub-sampled data. In~Figure~\ref{fig:Isabel_sampling_example}a a point rendering of the sampled data produced by the random sampling algorithm is shown and Figure~\ref{fig:Isabel_sampling_example}b depicts the point samples picked by the proposed association driven sampling algorithm. It can be observed that, since given the variables Pressure and Velocity, the~eyewall region of hurricane Isabel data set has a higher statistical association (as observed in the PMI field in Figure~\ref{fig:PMI_example}b), the~proposed sampling algorithm selects data points densely from such multivariate feature regions to preserve the feature more accurately. Another example of the proposed sampling algorithm applied to three variables (QGraup, QCloud, and~Precipitation) of Hurricane Isabel data is shown in Figure~\ref{fig:isabel_three_var_example}.  Figure~\ref{fig:isabel_three_var_example}a--c provide the visualization of the raw data for the three variables respectively and in Figure~\ref{fig:isabel_three_var_example}d we show the sample points selected by the proposed multivariate sampling algorithm. Note that, since this example uses three variables together, we have used the generalized pointwise information measure, specific correlation presented in  Equation~(\ref{eq:pmi_multivar}) for computing the multivariate statistical association for the data points considering three variables. It can be seen that the cloud and rain bands show strong statistical association and as a result, more data points are selected from such regions (the dark black regions in Figure~\ref{fig:isabel_three_var_example}d).
\vspace{-6pt}
\unskip
\section{Results}

\begin{figure}[H]
\centering
\begin{subfigure}[t]{0.23\linewidth}
     \centering
    \includegraphics[width=\linewidth]{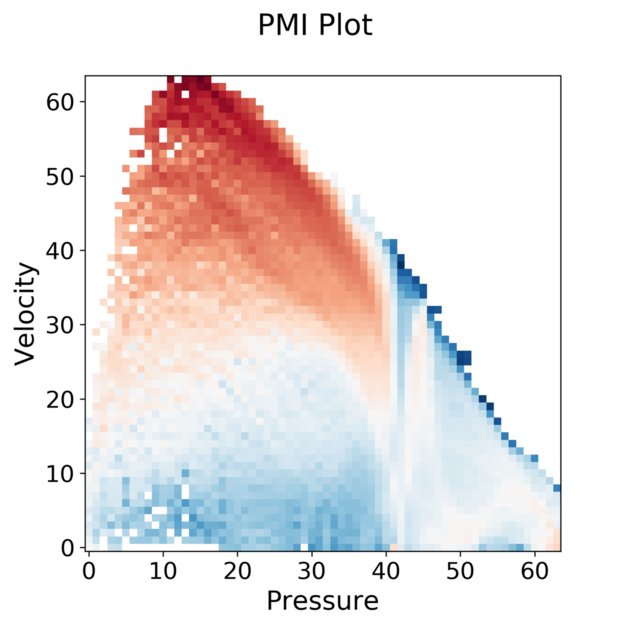}
    \caption{PMI plot with bin = 64.}
    \label{fig:Isabel_pmi_plot_P_VEL_64}
\end{subfigure}
~
\begin{subfigure}[t]{0.23\linewidth}
     \centering
    \includegraphics[width=\linewidth]{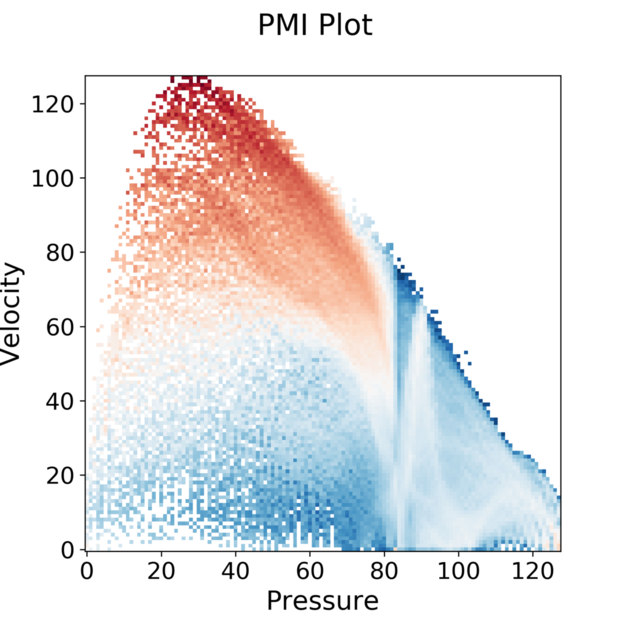}
    \caption{PMI plot with bin = 128.}
    \label{fig:Isabel_pmi_plot_P_VEL_128}
\end{subfigure}
~
\begin{subfigure}[t]{0.23\linewidth}
     \centering
    \includegraphics[width=\linewidth]{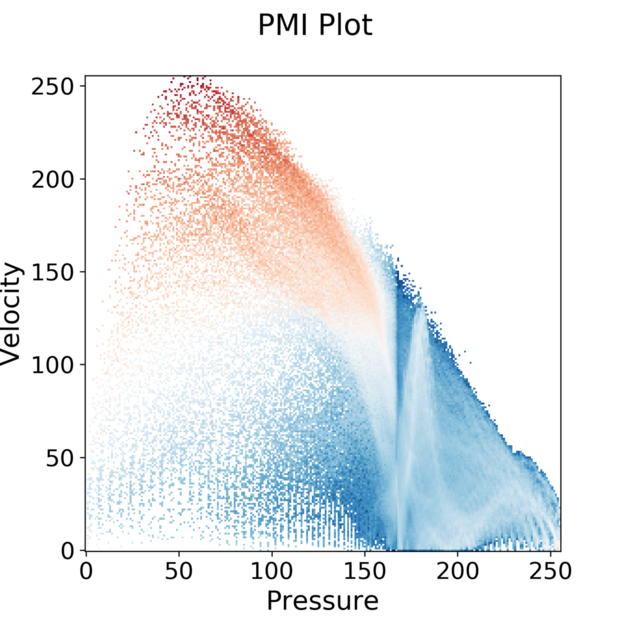}
    \caption{PMI plot with bin = 256.}
    \label{fig:Isabel_pmi_plot_P_VEL_256}
\end{subfigure}
\linebreak
\linebreak
\linebreak
\begin{subfigure}[t]{0.23\linewidth}
     \centering
    \includegraphics[width=\linewidth]{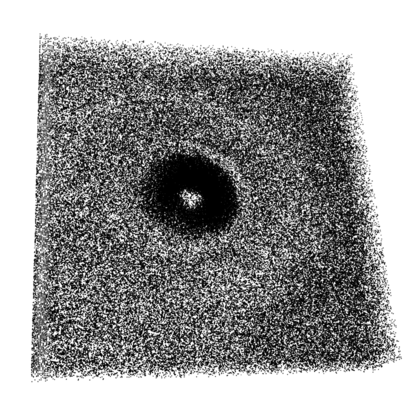}
    \caption{Samples selected with bin = 64.}
    \label{fig:samples_isabel_P_VEL_bin_64}
\end{subfigure}
~
\begin{subfigure}[t]{0.23\linewidth}
     \centering
    \includegraphics[width=\linewidth]{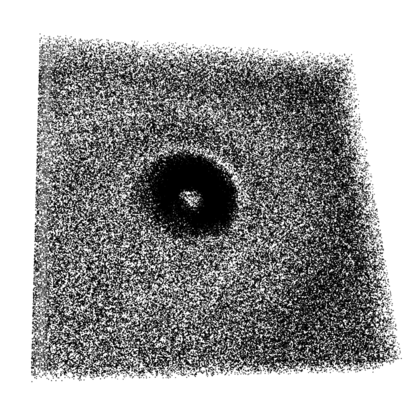}
    \caption{Samples selected with bin = 128.}
    \label{fig:samples_isabel_P_VEL_bin_128}
\end{subfigure}
~
\begin{subfigure}[t]{0.23\linewidth}
     \centering
    \includegraphics[width=\linewidth]{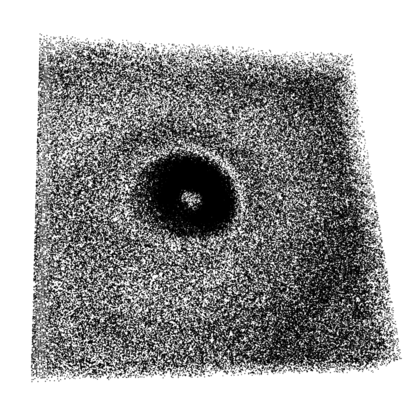}
    \caption{Samples selected with bin = 256.}
    \label{fig:samples_isabel_P_VEL_bin_256}
\end{subfigure}
\caption{Results of the proposed sampling technique when the number of histogram bins is varied while computing the information theoretic measure PMI. It is observed that the overall result remains similar without impacting the outcome of the sampling algorithm~significantly.}
\label{fig:pmi_plot_bin_study}
\end{figure}

In this section, we present the results of the case studies using several scientific simulation data sets to show the effectiveness of the proposed sampling algorithm. To~demonstrate the applicability of the sub-sampled data set, in~this work, we employ three important analysis tasks: (1) multivariate query-driven analysis, (2) reconstruction-based visualization of sampled data sets, and~(3) Multivariate correlation analysis. \bmark{While computing the multivariate histograms we studied the impact of a number of bins by varying it from $64$ to $256$. As~the number of bins increases, the~multivariate histogram becomes more refined as can be seen from Figure~\ref{fig:pmi_plot_bin_study}a--c. In~Figure~\ref{fig:pmi_plot_bin_study}, we show the results when different bin numbers are used for Isabel data set considering variables Pressure and Velocity. As~can be seen that the pattern in the PMI plots changes only slightly, but~the overall pattern remains similar. Also,~from~Figure~\ref{fig:pmi_plot_bin_study}d--f it is observed that the changes in the number of bins do not impact the proposed sampling algorithm significantly. Therefore, to~maintain consistency, for~all the experiments presented in this paper, we set the bin number to $128$. The~detailed results of the case studies are discussed below.}
\unskip


\subsection{Sample-Based Multivariate Query-Driven Visual~Analysis}
Query-driven visualization (QDV) is a well-known technique to understand and discover multivariate features in scientific datasets~\cite{gosink_query, query_driven, Hazarika_copula_multivar}. QDV is effective because it can reduce the computation workload significantly and help the domain experts to focus on the regions of interest. In~this work, to~help scientists to perform efficient multivariate query-driven feature exploration and visual analysis, we employ QDV to the sampled data sets directly. By~performing a multivariate query on the sampled data set, experts can quickly analyze and visualize the important features in the data set. After~analyzing the results of one specific query, the~experts can refine the query for performing a more detailed exploration.  However, note that, since the QDV is applied to the sampled data, the~result we obtain is an approximation. Given a user-specified query, e.g.,~the value of variable \mbox{$P$ $\geqslant$~x AND $Q$ $\leqslant$~y}, we first extract the data points that satisfy this query. To~visualize the results interactively, we~use point rendering techniques to directly visualize the query results in the spatial domain. \bmark{It is to be noted that, since the proposed sampling method prioritizes the data samples based on their PMI values, which reflects the existence of the multivariate association, the~proposed technique aims at primarily preserving such statistically associated regions with a higher precession compared to the regions in the data set where the statistical association is weak. Below~we present results of QDV from several multivariate scientific data sets and demonstrate the effectiveness of the proposed method.}

\subsubsection{Hurricane Isabel~Data}
We first apply the QDV analysis to Hurricane Isabel data set. Hurricane Isabel data set is used to study the impact of the hurricane Isabel on the coastal regions of the United States. The~dataset is courtesy of NCAR and the U.S. National Science Foundation (NSF) and was created using the Weather Research and Forecast (WRF) model. The~spatial resolution of the data is $250 \times 250 \times 50$. The~data set has $13$ variables and we have used Pressure and Velocity field for this study. To~understand the strength of the storm, scientists often explore the moderate to high-velocity regions. Furthermore, the~low-Pressure region at the eye of the hurricane also is an indicating factor for the strength of the storm~\cite{Hazarika_copula_multivar, gosink_1}. Therefore, we perform the query: $-$100 < Pressure < $-$4900 AND Velocity > 10 to analyze the low Pressure and moderate to high-Velocity regions in the sampled data~set. 

\begin{figure}[H]
\centering
\begin{subfigure}[H]{0.26\linewidth}
    \centering
    \includegraphics[width=\linewidth]{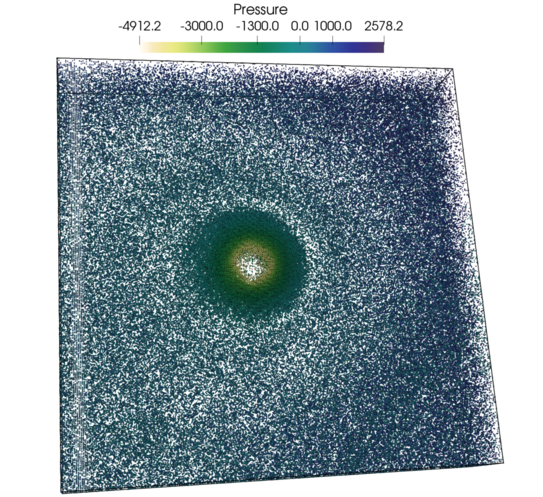}
    \caption{All the data points sampled by the proposed sampling~method.}
    \label{fig:pmi_sampled_7_P_VEL}
\end{subfigure}
~
\begin{subfigure}[H]{0.23\linewidth}
     \centering
    \includegraphics[width=\linewidth]{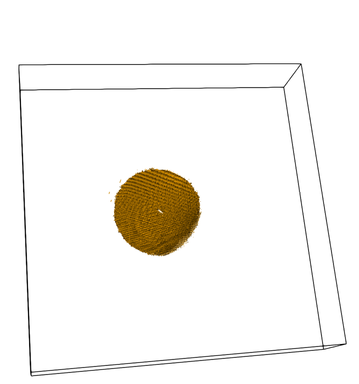}
    \caption{Result of the query when applied on raw~data.}
    \label{fig:query_raw_P_VEL}
\end{subfigure}
~
\begin{subfigure}[H]{0.23\linewidth}
    \centering
    \includegraphics[width=\linewidth]{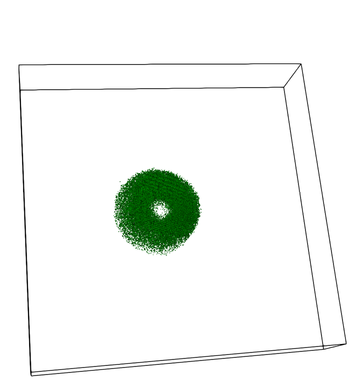}
    \caption{Result of the query when applied on PMI-based sampled~data.}
    \label{fig:query_pmi_7_P_VEL}
\end{subfigure}
~
\begin{subfigure}[H]{0.23\linewidth}
    \centering
    \includegraphics[width=\linewidth]{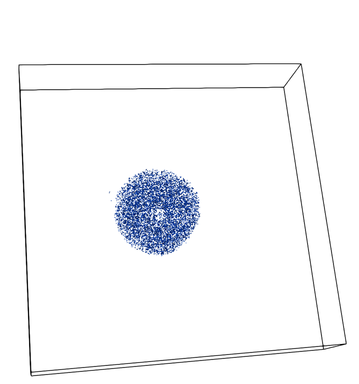}
    \caption{Result of the query applied to randomly sampled~data.}
    \label{fig:query_random_7_P_VEL}
\end{subfigure}
\caption{ {Visualization} 
 of multivariate query-driven analysis performed on the sampled data using Hurricane Isabel data set. The~multivariate query $-$100 < Pressure < $-$4900 AND Velocity > 10 is applied on the sampled data sets. Figure~\ref{fig:isabel_query}a shows all the points selected by the proposed sampling algorithm by using Pressure and Velocity variable. Figure~\ref{fig:isabel_query}b shows the data points selected by the query when applied to raw data. Figure~\ref{fig:isabel_query}c shows the points selected when the query is performed on the sub-sampled data produced by the proposed sampling scheme and Figure~\ref{fig:isabel_query}d presents the result of the query when applied to a randomly sampled data set. The~sampling fraction used in this experiment is $0.07$. }
\label{fig:isabel_query}
\end{figure}

In Figure~\ref{fig:isabel_query} the results of the above query is presented. Figure~\ref{fig:isabel_query}a shows all the sample points (sampling fraction $0.07$) that are selected initially before the query by applying the proposed information-driven sampling algorithm. The~points are colored using Pressure variable. As~can be seen that the information-driven sampling algorithm sampled points more densely where Pressure and Velocity show a strong statistical association. When the above multivariate query is applied on the original raw data, it returns a set of points shown in Figure~\ref{fig:isabel_query}b which is considered as the ground truth. Figure~\ref{fig:isabel_query}c provides the result of the same query when it is performed on the sub-sampled data (sampling fraction $0.07$) generated using our proposed algorithm, and~in Figure~\ref{fig:isabel_query}d we show the result of the query when applied on a randomly sampled data set (sampling fraction $0.07$). It can be observed from Figure~\ref{fig:isabel_query}c,d that the proposed sampling algorithm is able to answer the query more accurately compared to the randomly sampled data set since the proposed method returns a denser set of points close to the true result shown in Figure~\ref{fig:isabel_query}b. Since the random sampling method does not consider relationships among variables while selecting data points, it is unable to preserve such important features with high fidelity and results in a sparse set of points as can be seen in Figure~\ref{fig:isabel_query}d. 
\unskip

\subsubsection{Turbulent Combustion~Data}
Next, the~QDV analysis is applied to a Turbulent Combustion data set. This dataset is a turbulent simulation and consists of $5$ variables. Among~them, the~mixfrac variable is an important variable and denotes the proportion of fuel and oxidizer mass and this value generally provides the location of the flame where the chemical reaction rate exceeds the turbulent mixing rate~\cite{akiba_combustion, vis13_biswas}. The~spatial resolution of this dataset is $480\times720\times120$. Another important variable in this simulation is the mass fraction of the hydroxyl radical denoted as Y\_OH. Since it was previously observed that the mixfrac values around $0.42$ represent the flame region and Y\_OH field is nonuniform in the flame region~\cite{akiba_combustion, Hazarika_copula_multivar}, we~performed a multivariate query 0.3 < mixfrac < 0.7 AND 0.0006 < Y\_OH < 0.1 on the sampled data~set.

\begin{figure}[H]
\centering
\begin{subfigure}[H]{0.24\linewidth}
     \centering
    \includegraphics[width=\linewidth]{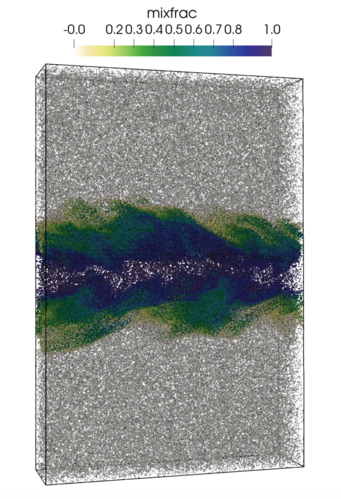}
    \caption{All the data points sampled by the proposed sampling~method.}
    \label{fig:pmi_sampled_7_mixfrac_Y_OH}
\end{subfigure}
~
\begin{subfigure}[H]{0.23\linewidth}
     \centering
    \includegraphics[width=\linewidth]{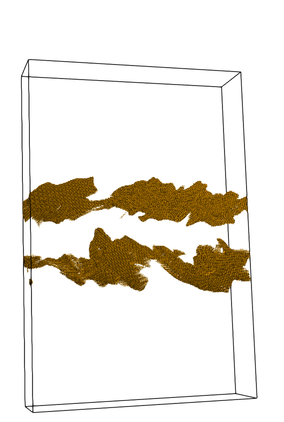}
    \caption{Result of the query when applied on raw~data.}
    \label{fig:query_raw_combustion}
\end{subfigure}
~
\begin{subfigure}[H]{0.23\linewidth}
     \centering
    \includegraphics[width=\linewidth]{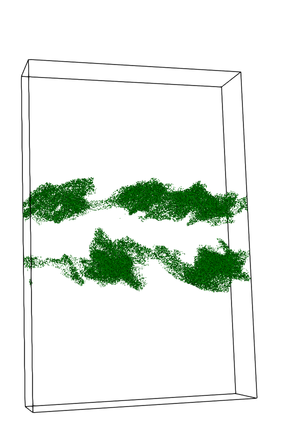}
    \caption{Result of the query when applied on PMI-based sampled~data.}
    \label{fig:query_pmi_7_mixfrac_Y_OH}
\end{subfigure}
~
\begin{subfigure}[H]{0.23\linewidth}
     \centering
    \includegraphics[width=\linewidth]{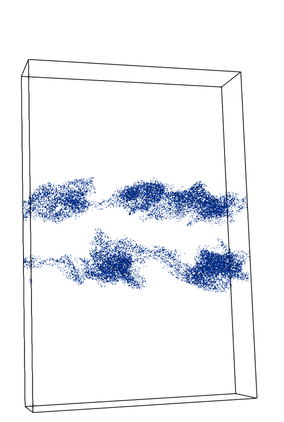}
    \caption{Result of the query applied to randomly sampled~data.}
    \label{fig:query_random_7_mixfrac_Y_OH}
\end{subfigure}
\caption{ {Visualization} of multivariate query-driven analysis performed on the sampled data using Turbulent Combustion data set. The~multivariate query 0.3 < mixfrac < 0.7 AND 0.0006 < Y\_OH 0.1 is applied on the sampled data sets. Figure~\ref{fig:combustion_query}a shows all the points selected by the proposed sampling algorithm by using mixfrac and Y\_OH variable. Figure~\ref{fig:combustion_query}b shows the data points selected by the query when applied to raw data. Figure~\ref{fig:combustion_query}c shows the points selected when the query is performed on the sub-sampled data produced by the proposed sampling scheme and Figure~\ref{fig:combustion_query}d presents the result of the query when applied to a randomly sampled data set. The~sampling fraction used in this experiment is $0.07$.}
\label{fig:combustion_query}
\end{figure}

Figure~\ref{fig:combustion_query}a shows the initial set of points selected by the proposed sampling method when a sampling fraction of $0.07$ is used. The~result of the above query when performed on the raw data is shown in Figure~\ref{fig:combustion_query}b. Figure~\ref{fig:combustion_query}c,d presents the results of the same query when performed on the sub-sampled data produced by the proposed method and on the randomly sampled data. For~this experiment, the~value of the sampling fraction is set to $0.07$. As~can be seen, the~result obtained from the samples generated by the proposed sampling method (Figure \ref{fig:combustion_query}c) generates higher quality compared to the result obtained from the randomly sampled~data.
\vspace{-6pt}
\unskip

\subsubsection{Asteroid Impact~Data}
Finally, we apply the query-driven analysis to the Asteroid impact data set. The~Deep Water Impact dataset~\cite{patchett2017deep} represents an ensemble of simulations run at the Los Alamos National Laboratory to study Asteroid Generated Tsunami or AGT. To~evaluate our sampling scheme, we have used one of the ensemble members in this work where the spatial resolution of the data is $300 \times 300 \times 300$. We~have used the volume fraction of water variable, denoted by v02 and the temperature variable denoted by tev to perform sub-sampling of the data. The~value of v02 lies between 0.0 and 1.0, where 1.0 means pure water. The~goal is to study the phenomena of ablation and ejecta material as the asteroid enters and subsequently impacts the water, sending a plume of material into the surrounding area and up into the atmosphere~\cite{gisler2018three}. Since the temperature increases as the asteroid enters the atmosphere, in~this work, we perform the following query to study the interaction between tev and v02: 0.13 < tev < 0.5 AND 0.45 < v02 < 1.0. 

\begin{figure}[H]
\centering
\begin{subfigure}[t]{0.4\linewidth}
     \centering
    \includegraphics[width=\linewidth]{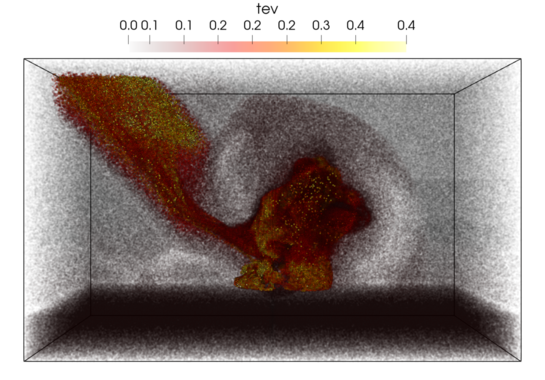}
    \caption{All the data points sampled by the proposed sampling~method.}
    \label{fig:pmi_sampled_7_tev_v02}
\end{subfigure}
~
\begin{subfigure}[t]{0.4\linewidth}
     \centering
    \includegraphics[width=\linewidth]{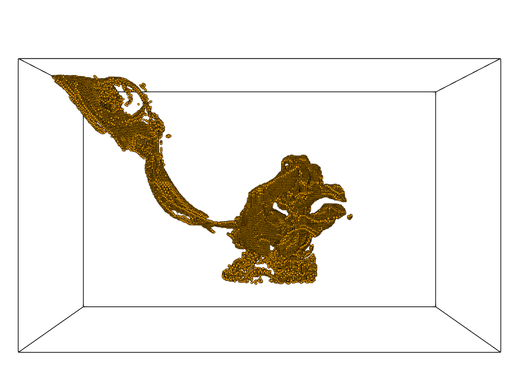}
    \caption{Result of the query when applied on raw~data.}
    \label{fig:query_raw_asteroid}
\end{subfigure}
\\
\vspace{2mm}
\begin{subfigure}[t]{0.4\linewidth}
     \centering
    \includegraphics[width=\linewidth]{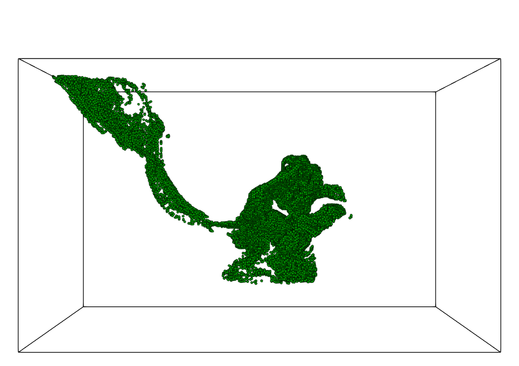}
    \caption{Result of the query when applied on PMI-based sampled~data.}
    \label{fig:query_pmi_7_tev_v02}
\end{subfigure}
~
\begin{subfigure}[t]{0.4\linewidth}
     \centering
    \includegraphics[width=\linewidth]{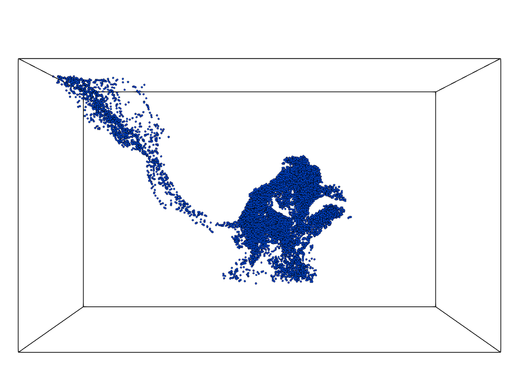}
    \caption{Result of the query applied on randomly sampled~data.}
    \label{fig:query_random_7_tev_v02}
\end{subfigure}
\caption{ {Visualization} of multivariate query driven analysis performed on the sampled data using Asteroid impact data set. The~multivariate query 0.13 < tev < 0.5 AND 0.45 < v02 1.0 is applied on the sampled data sets. Figure~\ref{fig:asteroid_query}a shows all the points selected by the proposed sampling algorithm by using tev and v02 variable. Figure~\ref{fig:asteroid_query}b shows the data points selected by the query when applied to raw data. Figure~\ref{fig:asteroid_query}c shows the points selected when the query is performed on the sub-sampled data produced by the proposed sampling scheme and Figure~\ref{fig:asteroid_query}d presents the result of the query when applied to a randomly sampled data set. The~sampling fraction used in this experiment is $0.07$.}
\label{fig:asteroid_query}
\end{figure}

The results of the query-driven analysis are presented in Figure~\ref{fig:asteroid_query}. Figure~\ref{fig:asteroid_query}a shows the point rendering of all the sub-sampled data points selected by the proposed sampling algorithm when a sampling fraction of $0.07$ is used. It can be seen that by analyzing the statistical association between tev and v02, the~proposed method selects samples more densely from the regions where the two variables show a high statistical association. The~query result obtained from the sub-sampled data produced by the proposed method (Figure \ref{fig:asteroid_query}c) is similar to the true query result produced from the raw data shown in Figure~\ref{fig:asteroid_query}b. It can be also seen that the query result performed on the randomly sampled data is sparse and misses some important features which are preserved well in the results obtained by the proposed sampling~method.

\unskip

\subsubsection{Quantitative Evaluation of Query-Driven~Analysis}
To study the accuracy of the query-driven analysis quantitatively, in~this work, we have used the similarity measure \textit{Jaccard index} \cite{jaccard}. The~Jaccard similarity index has been used extensively in the research community to measure the similarity between sample sets. Essentially, it measures the similarity between two finite sets and can be estimated as below:
\begin{equation}
 J(P,Q) = \frac{\left| P \cap Q\right|}{\left| P \cup Q\right|}  \label{eq:jaccard_index}
\end{equation} 
where $P$ and $Q$ are two finite sets. The~value of $J(P, Q)$ lies between 0 and 1, with~1 means \emph{P} and \emph{Q} are identical. In~our study of the query-driven analysis, set $P$ is the point set resulted from the query applied on the raw data and $Q$ is the point set resulted from either the proposed method or from the query applied on the randomly sampled data. Higher values of Jaccard index indicate more accurate query results. \bmark{In Table~\ref{qdv_table_isabel}, we provide the accuracy results of a range of queries that have been performed on Isabel data set using different variable combinations. As~can be seen that the proposed method performs well for all the queries compared to random sampling method. Note that, the~query presented in the second row of the Table~\ref{qdv_table_isabel}, where the query involves Pressure value range outside of the hurricane eye region (i.e., the~negative Pressure values), the~accuracy of the random sampling method is very similar to the proposed information-driven sampling method. This is due to the fact that in such regions, the~proposed sampling method did not pick samples densely due to the existence of a weak statistical association between Pressure and Velocity. Therefore, in~such weakly associated regions, random sampling based QDV can result in similar accuracy to that of the proposed method. Table~\ref{qdv_table_isabel} also shows QDV results for Pressure and QVapor variables of Isabel data set. As~can be seen that, for~a different range of Pressure and QVapor values, the~proposed method performs well. In~Tables~\ref{qdv_table_comb} and \ref{qdv_table_aste}, we provide the results of several QDV for Turbulent Combustion and Asteroid data sets respectively. Similar to the Isabel data set, it can be seen that the proposed method performs well for each of these data sets compared to the random sampling method for a range of values for different~variables.}


\subsection{Reconstruction-Based Visualization of Sampled~Data}
Besides query-driven analysis, to~visualize the data at its entirety at full resolution, we also perform reconstruction of the data from the sub-sampled data points. In~this work, we have adopted a linear interpolation based technique for data reconstruction such that the features in the data can be effectively visualized using volume visualization techniques. To~linearly interpolate the data, first, a~3D convex hull is generated using all the sampled points. Next, the~points are converted to a polygonal mesh by using Delaunay triangulation. Finally, for~each grid point in the reconstruction grid, the~value is calculated by linearly interpolating scalar values from the vertices of the simplex that encloses the current grid point. After~reconstruction, the~traditional raycasting-based volume rendering algorithm is used to visualize the features in the data set. In~the following, we first present the visualization of the reconstructed data for several multivariate data sets and then provide a quantitative image-based study of the visual quality of the results to demonstrate the efficacy of the proposed~method.
\begin{table}[H]
\centering
\tiny
\caption{Accuracy study of multivariate query-driven analysis for Isabel~data.}
\label{qdv_table_isabel}
\scalebox{0.73}
{
\begin{tabular}{ccccccccccc}
\toprule
\multirow{2}{*}{} & \multicolumn{2}{c}{\textbf{samp. frac: 0.01}} & \multicolumn{2}{c}{\textbf{samp. frac: 0.03}} & \multicolumn{2}{c}{\textbf{samp. frac: 0.05}} & \multicolumn{2}{c}{\textbf{samp. frac: 0.07}} & \multicolumn{2}{c}{\textbf{samp. frac: 0.09}} \\ \cmidrule{2-11} 
 & \textbf{Random} & \textbf{Proposed} & \textbf{Random} & \textbf{Proposed} & \textbf{Random} & \textbf{Proposed} & \textbf{Random} & \textbf{Proposed} & \textbf{Random} & \textbf{Proposed} \\ \midrule
\begin{tabular}[c]{@{}c@{}}Isabel data\\ ($-$100 \textless\, Pres \textless\, $-$4900\\ \& Vel \textgreater\, 10)\end{tabular} & 0.0096 & 0.0468 & 0.029 & 0.143 & 0.048 & 0.233 & 0.0676 & 0.315 & 0.0846 & 0.388 \\ \midrule
\begin{tabular}[c]{@{}c@{}}Isabel data\\ (0 \textless\, Pres \textless\, 1500\\ \& 10 \textless\, Vel \textless\, 35)\end{tabular} & 0.0116 & 0.0103 & 0.0293 & 0.0332 & 0.05 & 0.0524 & 0.0724 & 0.078 & 0.0842 & 0.0969 \\ \midrule
\begin{tabular}[c]{@{}c@{}}Isabel data\\ ($-$100 \textless\, Pres \textless\, $-$4900\\ \& Qva \textgreater\, 0.017)\end{tabular} & 0.0086 & 0.0912 & 0.033 & 0.163 & 0.05 & 0.266 & 0.0637 & 0.284 & 0.086 & 0.314 \\ \midrule
\begin{tabular}[c]{@{}c@{}}Isabel data\\ (Pres \textgreater\, 300\\ \& 0.02 \textless\, Qva \textless\, 0.03)\end{tabular} & 0.0088 & 0.023 & 0.0159 & 0.0585 & 0.062 & 0.1241 & 0.0726 & 0.1507 & 0.0975 & 0.2446 \\ \midrule
\end{tabular}
}
\end{table}
\unskip

\begin{table}[H]
\centering
\tiny
\caption{Accuracy study of multivariate query-driven analysis for Combustion~data.}
\label{qdv_table_comb}
\scalebox{0.71}
{
\begin{tabular}{ccccccccccc}
\toprule
\multirow{2}{*}{} & \multicolumn{2}{c}{\textbf{samp. frac: 0.01}} & \multicolumn{2}{c}{\textbf{samp. frac: 0.03}} & \multicolumn{2}{c}{\textbf{samp. frac: 0.05}} & \multicolumn{2}{c}{\textbf{samp. frac: 0.07}} & \multicolumn{2}{c}{\textbf{samp. frac: 0.09}} \\ \cmidrule{2-11} 
 & \textbf{Random} & \textbf{Proposed} & \textbf{Random} & \textbf{Proposed} & \textbf{Random} & \textbf{Proposed} & \textbf{Random} & \textbf{Proposed} & \textbf{Random} & \textbf{Proposed} \\ \midrule
\begin{tabular}[c]{@{}c@{}}Combustion data\\ (0.3 \textless\, mixfrac \textless\, 0.7 \\ \& 0.0006 \textless\, Y\_OH \textless\, 0.1)\end{tabular} & 0.0099 & 0.0275 & 0.029 & 0.081 & 0.048 & 0.135 & 0.0671 & 0.191 & 0.0862 & 0.244 \\ \midrule
\begin{tabular}[c]{@{}c@{}}Combustion data\\ (0.7 \textless\, mixfrac \textless\, 1.0 \\ \& 0.0005 \textless\, Y\_OH \textless\, 0.0019)\end{tabular} & 0.00884 & 0.0329 & 0.0291 & 0.1139 & 0.0474 & 0.1892 & 0.0686 & 0.2636 & 0.0877 & 0.3518 \\ \midrule
\end{tabular}
}
\end{table}
\unskip

\begin{table}[H]
\centering
\tiny
\caption{Accuracy study of multivariate query-driven analysis for Asteroid~data.}
\label{qdv_table_aste}
\scalebox{0.75}
{
\begin{tabular}{ccccccccccc}
\toprule
\multirow{2}{*}{} & \multicolumn{2}{c}{\textbf{samp. frac: 0.01}} & \multicolumn{2}{c}{\textbf{samp. frac: 0.03}} & \multicolumn{2}{c}{\textbf{samp. frac: 0.05}} & \multicolumn{2}{c}{\textbf{samp. frac: 0.07}} & \multicolumn{2}{c}{\textbf{samp. frac: 0.09}} \\ \cmidrule{2-11} 
 & \textbf{Random} & \textbf{Proposed} & \textbf{Random} & \textbf{Proposed} & \textbf{Random} & \textbf{Proposed} & \textbf{Random} & \textbf{Proposed} & \textbf{Random} & \textbf{Proposed} \\ \midrule
\begin{tabular}[c]{@{}c@{}}Asteroid data\\ (0.13 \textless\, tev \textless\, 0.5 \\ \& 0.45 \textless\, v02 \textless\, 1.0)\end{tabular} & 0.013 & 0.067 & 0.029 & 0.202 & 0.0479 & 0.328 & 0.0678 & 0.431 & 0.086 & 0.52 \\ \midrule
\begin{tabular}[c]{@{}c@{}}Asteroid data\\ (0.1 \textless\, tev \textless\, 0.3\\ \& 0.01 \textless\, v02 \textless\, 0.6)\end{tabular} & 0.0097 & 0.0827 & 0.0302 & 0.2497 & 0.0491 & 0.4154 & 0.0668 & 0.5777 & 0.0866 & 0.7083 \\ \midrule
\end{tabular}
}
\end{table}
\unskip

\subsubsection{Hurricane Isabel~Data}

\begin{figure}[H]
\centering
\begin{subfigure}[t]{0.25\linewidth}
     \centering
    \framebox{\includegraphics[width=\linewidth]{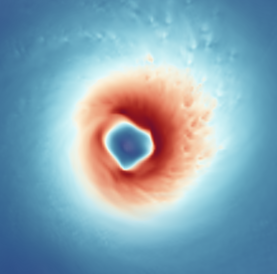}}
    \caption{Visualization using raw~data.}
    \label{fig:VEL_raw}
\end{subfigure}
~~~
\begin{subfigure}[t]{0.25\linewidth}
     \centering
    \framebox{\includegraphics[width=\linewidth]{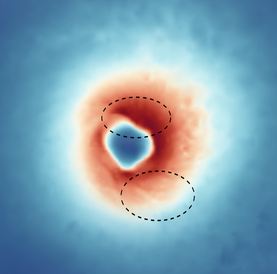}}
    \caption{Reconstructed data visualization using proposed~sampling algorithm.}
    \label{fig:VEL_pmi_5}
\end{subfigure}
~~~
\begin{subfigure}[t]{0.25\linewidth}
     \centering
    \framebox{\includegraphics[width=\linewidth]{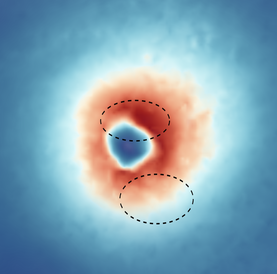}}
    \caption{Reconstructed data visualization using random sampling~algorithm.}
    \label{fig:VEL_random_5}
\end{subfigure}
\caption{Reconstruction-based visualization of Velocity field of Hurricane Isabel data set. Linear interpolation is used to reconstruct the data from the sub-sampled data sets. Figure~\ref{fig:isabel_recon_VEL}a shows the result from the original raw data. Figure~\ref{fig:isabel_recon_VEL}b provides the reconstruction result from the sub-sampled data generated by the proposed method, and~Figure~\ref{fig:isabel_recon_VEL}c presents the result of reconstruction from random sampled data. The~sampling fraction used in this experiment is $0.05$.}
\label{fig:isabel_recon_VEL}
\end{figure}

Figure~\ref{fig:isabel_recon_VEL} shows the reconstructed data visualization of Velocity field for Hurricane Isabel data set. \bmark{The visualization is focused on the feature in the data set, i.e.,~the hurricane eye region. The~eye of the hurricane is an important feature where strong wind exists indicated by the dark reddish regions in the image. At~the core of the eye, the~velocity is low as seen from the blue color and around it the eyewall is generally the region where the destructive winds exist.} In Figure~\ref{fig:isabel_recon_VEL}, by~comparing the Figure~\ref{fig:isabel_recon_VEL}a, which shows the visualization of the Velocity field generated using the full resolution raw data, we can observe that the visualization produced by the proposed sampling technique (Figure \ref{fig:isabel_recon_VEL}b) matches quite well to the raw data image. \bmark{However, it can be seen that the image produced from randomly sampled data presented in Figure~\ref{fig:isabel_recon_VEL}c is missing some fine details around the hurricane eye region as shown by the dotted black lines in the images which are preserved more accurately by the proposed sampling algorithm.} 
\subsubsection{Turbulent Combustion~Data}

\begin{figure}[H]
\centering
\begin{subfigure}[t]{0.25\linewidth}
     \centering
    \framebox{\includegraphics[width=\linewidth, height=1.8in]{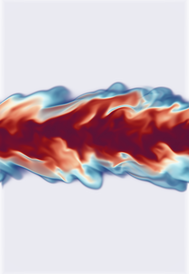}}
    \caption{Visualization using raw~data.}
    \label{fig:mixfrac_raw}
\end{subfigure}
~~~
\begin{subfigure}[t]{0.25\linewidth}
     \centering
    \framebox{\includegraphics[width=\linewidth, height=1.8in]{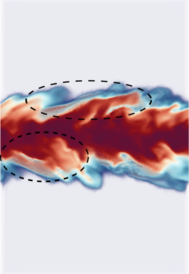}}
    \caption{Reconstructed data visualization using proposed~sampling algorithm.}
    \label{fig:mixfrac_pmi_5}
\end{subfigure}
~~~
\begin{subfigure}[t]{0.25\linewidth}
     \centering
    \framebox{\includegraphics[width=\linewidth, height=1.8in]{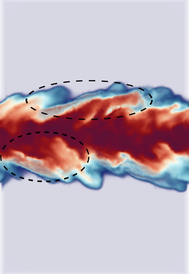}}
    \caption{Reconstructed data visualization using random sampling~algorithm.}
    \label{fig:mixfrac_random_5}
\end{subfigure}
\caption{Reconstruction-based visualization of mixfrac field of Turbulent Combustion data set. Linear interpolation is used to reconstruct the data from the sub-sampled data sets. Figure~\ref{fig:combustion_recon_mixfrac}a shows the result from the original raw data. Figure~\ref{fig:combustion_recon_mixfrac}b provides the reconstruction result from the sub-sampled data generated by the proposed method, and~Figure~\ref{fig:combustion_recon_mixfrac}c presents the result of reconstruction from random sampled data. The~sampling fraction used in this experiment is $0.05$.}
\label{fig:combustion_recon_mixfrac}
\end{figure}

\begin{figure}[H]
\centering
\begin{subfigure}[t]{0.25\linewidth}
     \centering
    \framebox{\includegraphics[width=\linewidth, height=1.8in]{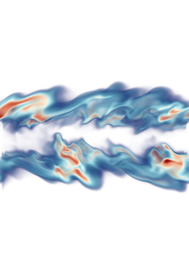}}
    \caption{Visualization using raw~data.}
    \label{fig:Y_OH_raw}
\end{subfigure}
~~~
\begin{subfigure}[t]{0.25\linewidth}
     \centering
    \framebox{\includegraphics[width=\linewidth, height=1.8in]{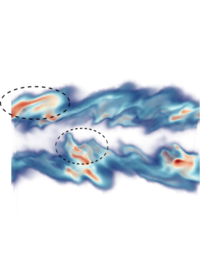}}
    \caption{Reconstructed data visualization using proposed sampling algorithm.}
    \label{fig:Y_OH_pmi_5}
\end{subfigure}
~~~
\begin{subfigure}[t]{0.25\linewidth}
     \centering
    \framebox{\includegraphics[width=\linewidth, height=1.8in]{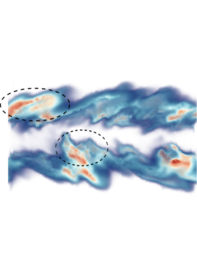}}
    \caption{Reconstructed data visualization using random sampling algorithm.}
    \label{fig:Y_OH_random_5}
\end{subfigure}
\caption{Reconstruction-based visualization of Y\_OH field of Turbulent Combustion data set. Linear interpolation is used to reconstruct the data from the sub-sampled data sets. Figure~\ref{fig:combustion_recon_YOH}a shows the result from the original raw data. Figure~\ref{fig:combustion_recon_YOH}b provides the reconstruction result from the sub-sampled data generated by the proposed method, and~Figure~\ref{fig:combustion_recon_YOH}c presents the result of reconstruction from random sampled data. The~sampling fraction used in this experiment is $0.05$.}
\label{fig:combustion_recon_YOH}
\end{figure}

The reconstruction based visualization for mixfrac and Y\_OH fields of Turbulent Combustion data is shown in Figures~\ref{fig:combustion_recon_mixfrac} and \ref{fig:combustion_recon_YOH} respectively. 
\bmark{In this case, the~visualization of the turbulent mixing region is considered as an important feature since the mixfrac and Y\_OH fields interact with each other in these regions. Scientists often try to find regions where the high Y\_OH values exist and how are the high Y\_OH values distributed on mixture fraction surfaces to study the flame structure~\cite{akiba_combustion}.} The sampling fraction used in this experiment is $0.05$ for all the sampling algorithms. For~the mixfrac field, both the proposed sampling technique (Figure \ref{fig:combustion_recon_mixfrac}b) and the random sampling technique (see Figure~\ref{fig:combustion_recon_mixfrac}c) produce images which are visually very similar to that of the raw data image provided in Figure~\ref{fig:combustion_recon_mixfrac}a. There are some minor differences in the images as highlighted by the black dotted lines. However, the~proposed multivariate association-based sampling technique produces more accurate visualization for the Y\_OH field compared to the random sampling algorithm. \bmark{By comparing Figure~\ref{fig:combustion_recon_YOH}b,c, it is observed that the regions where Y\_OH values are high (indicated by the black dotted lines) are more accurate in Figure~\ref{fig:combustion_recon_YOH}b obtained by our proposed sampling scheme. Furthermore, as~mentioned above, such high valued regions are important since the  scientists often study such high Y\_OH valued regions and study its relationship with the mixture fraction values~\cite{akiba_combustion}.}
\unskip

\subsubsection{Asteroid Impact~Data}

\begin{figure}[H]
\centering
\begin{subfigure}[t]{0.32\linewidth}
     \centering
    \includegraphics[width=\linewidth]{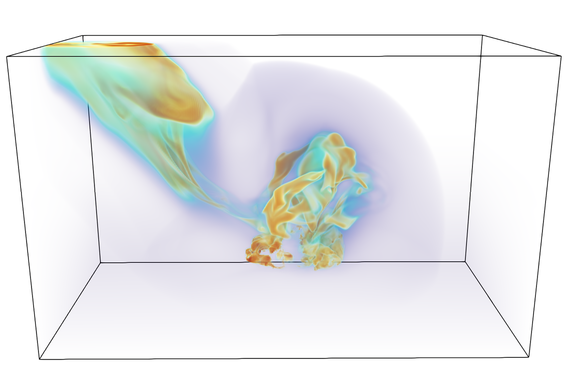}
    \caption{Visualization using raw~data.}
    \label{fig:tev_raw}
\end{subfigure}
~
\begin{subfigure}[t]{0.32\linewidth}
     \centering
    \includegraphics[width=\linewidth]{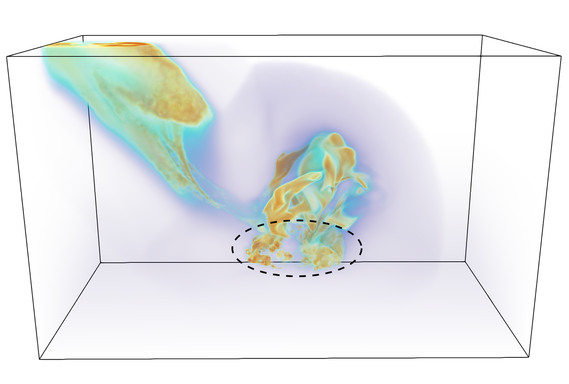}
    \caption{Reconstructed data visualization using proposed sampling algorithm.}
    \label{fig:tev_pmi_5}
\end{subfigure}
~
\begin{subfigure}[t]{0.32\linewidth}
     \centering
    \includegraphics[width=\linewidth]{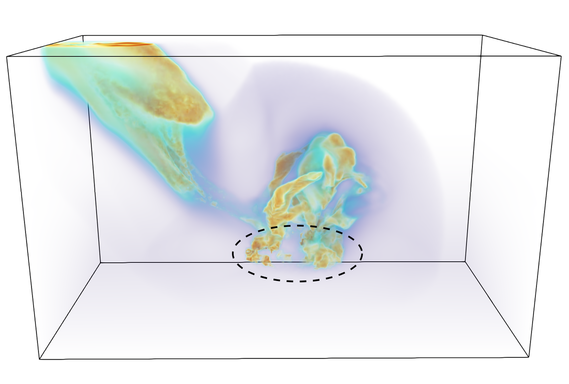}
    \caption{Reconstructed data visualization using random sampling~algorithm.}
    \label{fig:tev_random_5}
\end{subfigure}
\caption{Reconstruction-based visualization of tev field of Asteroid impact data set. Linear~interpolation is used to reconstruct the data from the sub-sampled data sets. Figure~\ref{fig:asteroid_recon_tev}a shows the result from the original raw data. Figure~\ref{fig:asteroid_recon_tev}b provides the reconstruction result from the sub-sampled data generated by the proposed method, and~Figure~\ref{fig:asteroid_recon_tev}c presents the result of reconstruction from random sampled data. The~sampling fraction used in this experiment is $0.05$.}
\label{fig:asteroid_recon_tev}
\end{figure}

Finally, we show the reconstruction-based visualizations of tev (temperature) field in Figure~\ref{fig:asteroid_recon_tev}. Sampling fraction of $0.05$ is used for this study. \bmark{In this visualization, the~dark orange regions indicate the places where the temperature value is high and is often indicative of the existence of the asteroid. This is due to the fact that as the asteroid enters the atmosphere, the~temperature increases due to the friction with the atmosphere. On~the lower right side, the~region shows the impact of the asteroid with the surface of the water.} Figure~\ref{fig:asteroid_recon_tev}b shows the reconstructed image when the sampled data from the proposed method is used and by comparing it with raw data image (Figure \ref{fig:asteroid_recon_tev}a), we can observe that the reconstructed image matches quite well to the raw data image. Figure~\ref{fig:asteroid_recon_tev}c shows the visualization produced using the randomly sampled data. It is seen that even though the image quality of the random sampling method is good, however, the~proposed method preserves the fine structures in the image better (as highlighted by the dotted black regions) compared to the image obtained by randomly sampled~data. 

\subsubsection{Image-Based Quantitative Evaluation of Reconstruction-Based~Visualization}
After the above qualitative analysis of reconstruction-based data visualization, in~this section, we provide a comprehensive study of the quality of the images generated by both the proposed sampling method and random sampling method. To~study the quality, we performed an image-based comparison of the results obtained from both the sampling technique to the ground truth generated from the full resolution raw data under varying sampling fractions. Note that, for~the generation of images, the~same rendering parameters are used for each data set. To~study the image quality quantitatively, we have used the Structural Similarity Index (SSIM) \cite{SSIM}. SSIM is a well-known image similarity measure frequently used in the Computer Vision community and measures the distortion between an image with respect to a reference image. This method exploits the fact that human visual perception is highly adapted for extracting structural information from an image and is designed to capture the degradation of structural information. SSIM compares the images based on the variation of luminance, contrast, and~structural information. Therefore, the~SSIM measure~\cite{SSIM} between two images $I_1$ and $I_2$ is represented as:
\begin{equation}
 SSIM(I_1,I_2) =  l(I_1,I_2)^{a} \cdot c(I_1,I_2)^{b} \cdot s(I_1,I_2)^{c}  \label{eq:ssim_index}
\end{equation}
where $l(I_1,I_2)$ represents luminance similarity, $c(I_1,I_2)$ denotes contrast similarity and $s(I_1,I_2)$ contains the structural information. In~the above equation, $a$, $b$, and~$c$ are three parameters used to assign the relative importance to the three corresponding components in the measure and the default value for those parameters are set to $1$ indicating equal importance. Using SSIM, the~comparison between the images are done on local patches first and finally the mean values of all SSIM values from all the patches are computed which represents the final SSIM between two images. Besides~SSIM, we have also used Mean Squared Error (MSE) as our second image quality comparison measure. Note that higher values of SSIM indicate better similarity between the test image and the reference image and since MSE estimates an error quantity, lower values are considered more accurate for this measure. In~our study, the~raw images are considered as the reference image and images produced by the proposed sampling algorithm and random sampling are compared against the reference~image.

The results of the image-based quality comparison for Hurricane Isabel data set is presented in Tables~\ref{isabel_P_VEL_table} and \ref{isabel_P_QVA_table}. Table~\ref{isabel_P_VEL_table} shows results when Pressure and Velocity fields are used for sampling, and~Table~\ref{isabel_P_QVA_table} depicts results when Pressure and QVapor variables are used. Table~\ref{combustion_mixfrac_Y_OH_table} summarizes the results of image-based comparison for Turbulent Combustion data where mixfrac and Y\_OH fields are used and Table~\ref{asteroid_tev_v02_table} shows results for Asteroid Impact data using tev and v02 field. From~all the results obtained by these three multivariate data sets, it can be observed that the proposed method has produced better image quality consistently for the similarity measures across different sampling fractions which demonstrates the superiority of the proposed multivariate association-driven sampling algorithm over the existing popular random sampling algorithm. We also observe that for v02 field of Asteroid data set (Table \ref{asteroid_tev_v02_table}), MSE measures are slightly better for random sampling algorithm, however, when the structural similarity is estimated through SSIM measure, the~proposed sampling algorithm produces more accurate results than random sampling technique. This indicates that the proposed sampling algorithm is able to preserve the structure of the features in the visualization more~accurately.

\begin{table}[H]
\centering
\caption{Image-based comparison of Pressure and Velocity field of Isabel data~set.}
\label{isabel_P_VEL_table}
\begin{tabular}{ccccccl}
\toprule
\multirow{2}{*}{\begin{tabular}[c]{@{}c@{}}\textbf{Isabel Pressure}\\ \textbf{Field}\end{tabular}} & \multicolumn{2}{c}{\textbf{samp. frac: 0.01}} & \multicolumn{2}{c}{\textbf{samp. frac: 0.03}} & \multicolumn{2}{c}{\textbf{samp. frac: 0.05}} \\ \cmidrule{2-7} 
 & \textbf{Random} & \textbf{Proposed} & \textbf{Random} & \textbf{Proposed} & \textbf{Random} & \textbf{Proposed} \\ \midrule
SSIM & 0.9844 & 0.9915 & 0.9916 & 0.9931 & 0.9926 & 0.9939 \\ \midrule
MSE & 6.5563 & 1.9267 & 2.5239 & 1.2559 & 2.0576 & 0.8961 \\
\midrule
\multirow{2}{*}{\begin{tabular}[c]{@{}c@{}}\textbf{Isabel Velocity}\\ \textbf{Field}\end{tabular}} & \multicolumn{2}{c}{\textbf{samp. frac: 0.01}} & \multicolumn{2}{c}{\textbf{samp. frac: 0.03}} & \multicolumn{2}{c}{\textbf{samp. frac: 0.05}} \\ \cmidrule{2-7} 
 & \textbf{Random} & \textbf{Proposed} & \textbf{Random} & \textbf{Proposed} & \textbf{Random} & \textbf{Proposed} \\ \midrule
SSIM & 0.9234 & 0.9559 & 0.9427 & 0.9649 & 0.9516 & 0.9702 \\ \midrule
MSE & 13.9638 & 8.492 & 10.6865 & 6.0452 & 8.1166 & 5.0213 \\ \midrule
\end{tabular}
\end{table}
\unskip

\begin{table}[H]
\centering
\caption{Image-based comparison of Pressure and QVapor field of Isabel data~set.}
\label{isabel_P_QVA_table}
\begin{tabular}{ccccccl}
\toprule
\multirow{2}{*}{\begin{tabular}[c]{@{}c@{}}\textbf{Isabel Pressure}\\ \textbf{Field}\end{tabular}} & \multicolumn{2}{c}{\textbf{samp. frac: 0.01}} & \multicolumn{2}{c}{\textbf{samp. frac: 0.03}} & \multicolumn{2}{c}{\textbf{samp. frac: 0.05}} \\ \cmidrule{2-7} 
 & \textbf{Random} & \textbf{Proposed} & \textbf{Random} & \textbf{Proposed} & \textbf{Random} & \textbf{Proposed} \\ \midrule
SSIM & 0.9834 & 0.9919 & 0.9915 & 0.9926 & 0.9916 & 0.9929 \\ \midrule
MSE & 6.5982 & 2.3903 & 3.0432 & 2.0518 & 3.0987 & 1.9561 \\ 
\midrule
\multirow{2}{*}{\begin{tabular}[c]{@{}c@{}}\textbf{Isabel QVapor}\\ \textbf{Field}\end{tabular}} & \multicolumn{2}{c}{\textbf{samp. frac: 0.01}} & \multicolumn{2}{c}{\textbf{samp. frac: 0.03}} & \multicolumn{2}{c}{\textbf{samp. frac: 0.05}} \\ \cmidrule{2-7} 
 & \textbf{Random} & \textbf{Proposed} & \textbf{Random} & \textbf{Proposed} & \textbf{Random} & \textbf{Proposed} \\ \midrule
SSIM & 0.7495 & 0.7726 & 0.7745 & 0.7899 & 0.7838 & 0.80521 \\ \midrule
MSE & 12.7532 & 11.8243 & 10.2122 & 9.2676 & 9.262 & 8.2 \\ \midrule
\end{tabular}
\end{table}
\unskip

\begin{table}[H]
\centering
\caption{Image-based comparison of Mixfrac and Y\_OH field of Combustion data~set.}
\label{combustion_mixfrac_Y_OH_table}
\begin{tabular}{ccccccc}
\toprule
\multirow{2}{*}{\begin{tabular}[c]{@{}c@{}}\textbf{Combustion mixfrac}\\ \textbf{Field}\end{tabular}} & \multicolumn{2}{c}{\textbf{samp. frac: 0.01}} & \multicolumn{2}{c}{\textbf{samp. frac: 0.03}} & \multicolumn{2}{c}{\textbf{samp. frac: 0.05}} \\ \cmidrule{2-7} 
 & \textbf{Random} & \textbf{Proposed} & \textbf{Random} & \textbf{Proposed} & \textbf{Random} & \textbf{Proposed} \\ \midrule
SSIM & 0.8913 & 0.9138 & 0.9373 & 0.9538 & 0.9452 & 0.9708 \\ \midrule
MSE & 14.5252 & 12.2813 & 9.376 & 7.9371 & 18.141 & 5.7753 \\
\midrule
\multirow{2}{*}{\begin{tabular}[c]{@{}c@{}}\textbf{Combustion Y\_OH}\\ \textbf{Field}\end{tabular}} & \multicolumn{2}{c}{\textbf{samp. frac: 0.01}} & \multicolumn{2}{c}{\textbf{samp. frac: 0.03}} & \multicolumn{2}{c}{\textbf{samp. frac: 0.05}} \\ \cmidrule{2-7} 
 & \textbf{Random} & \textbf{Proposed} & \textbf{Random} & \textbf{Proposed} & \textbf{Random} & \textbf{Proposed} \\ \midrule
SSIM & 0.8868 & 0.9061 & 0.9401 & 0.9565 & 0.955 & 0.9739 \\ \midrule
MSE & 14.4677 & 13.6179 & 9.111 & 8.0836 & 7.4155 & 5.9128 \\ \midrule
\end{tabular}
\end{table}
\unskip

\begin{table}[H]
\centering
\caption{Image-based comparison of tev and v02 field of Asteroid data~set.}
\label{asteroid_tev_v02_table}
\begin{tabular}{ccccccc}
\toprule
\multirow{2}{*}{\begin{tabular}[c]{@{}c@{}}\textbf{Asteroid te}v\\ \textbf{Field}\end{tabular}} & \multicolumn{2}{c}{\textbf{samp. frac: 0.01}} & \multicolumn{2}{c}{\textbf{samp. frac: 0.03}} & \multicolumn{2}{c}{\textbf{samp. frac: 0.05}} \\ \cmidrule{2-7} 
 & \textbf{Random} & \textbf{Proposed} & \textbf{Random} & \textbf{Proposed} & \textbf{Random} & \textbf{Proposed} \\ \midrule
SSIM & 0.9746 & 0.9813 & 0.9808 & 0.9885 & 0.9849 & 0.9908 \\ \midrule
MSE & 4.93 & 4.3499 & 3.8366 & 3.1976 & 3.2674 & 2.7139 \\ 
\midrule
\multirow{2}{*}{\begin{tabular}[c]{@{}c@{}}\textbf{Asteroid v02}\\ \textbf{Field}\end{tabular}} & \multicolumn{2}{c}{\textbf{samp. frac: 0.01}} & \multicolumn{2}{c}{\textbf{samp. frac: 0.03}} & \multicolumn{2}{c}{\textbf{samp. frac: 0.05}} \\ \cmidrule{2-7} 
 & \textbf{Random} & \textbf{Proposed} & \textbf{Random} & \textbf{Proposed} & \textbf{Random} & \textbf{Proposed} \\ \midrule
SSIM & 0.7898 & 0.8121 & 0.7972 & 0.8213 & 0.8064 & 0.8326 \\ \midrule
MSE & 31.27 & 32.91 & 26.301 & 27.656 & 23.9335 & 25.4177 \\ \midrule

\end{tabular}
\end{table}

\bmark{

\subsection{Multivariate Correlation Analysis of the Proposed Sampling~Method}
In this section, we evaluate the proposed method in preserving the multivariate association among selected variables. Since the proposed PMI-based sampling technique prioritizes data points based on their shared information, the~technique aims at preserving the relationship between variable combinations in the strongly associated regions. More specifically, the~proposed technique aims at recovering back the statistical association in the important regions where a strong local dependence among variables exists. Note that, as~shown above, these strongly statistically associated regions often indicate important scientific features in the data set where multiple variables interact with each other. For~example, in~Hurricane Isabel data set, we have shown that the low Pressure valued regions (\mbox{i.e., negative} Pressure regions) and moderate to high wind velocity regions show the important feature ``hurricane eye''. The~proposed technique in this work aims at preserving such regions with higher fidelity such that the multivariate dependence among variables in such regions can be recovered. Therefore, to~evaluate the quality of the preservation of statistical association, we have used two measures in this work. The~first measure is \textit{Pearson's correlation coefficient}, which measures the linear correlation between variables. In~addition, since the variables can have non-linear dependence, we have also used \textit{distance correlation} \cite{dist_corr} to estimate the non-linear correlation among variables using the reconstructed fields. While Pearson's correlation coefficient takes values between $-$1.0 and 1.0 where $-$1.0 indicates total negative correlation and 1.0 indicates the total positive correlation, distance correlation takes values only between 0 and 1.0 where 0 indicates~independence.

The first evaluation is done on the reconstructed data set by specifically focusing on the important feature regions of the data sets. In~Figure~\ref{fig:correlation_VOI}, we present the regions of interest (ROI) for each data set used in this study. The~black box shown in the images indicates the ROI where the correlation study is performed. Note that, for~Hurricane Isabel data, we have selected the Hurricane eye region (see Figure~\ref{fig:correlation_VOI}a), for~Combustion data, we have selected the turbulent flame region as our ROI (Figure~\ref{fig:correlation_VOI}b), and~in the Asteroid impact data set, we have used the region where the asteroid has impacted the sea surface and the water is ejected into the atmosphere (Figure \ref{fig:correlation_VOI}c). Table~\ref{corr_study} provides the different correlation values that we obtained from different data sets when the correlation measures are computed in the reconstructed fields considering only the feature regions indicated by the ROIs in Figure~\ref{fig:correlation_VOI}. As~can be seen from Table~\ref{corr_study}, the~proposed method is able to preserve both the linear and non-linear correlation better compared to the random sampling method. Next, we estimated these two correlation measures by considering the full reconstructed data. The~results are shown in Table~\ref{corr_study_full}. As~we can see that when the full reconstructed data is considered, for~Isabel and Combustion data set, the~proposed information-driven technique is performing better than the random sampling technique. However, the~random sampling technique achieves a more accurate result in the Asteroid data. These~results demonstrate that, when full data is considered, both these techniques are competitive. However, since the proposed method is designed to preserve the important statistically associated regions with higher fidelity, it is able to recover the linear and non-linear dependence between variables accurately compared to random~sampling.

\begin{figure}[H]
\centering
\begin{subfigure}[H]{0.27\linewidth}
     \centering
    \includegraphics[width=\linewidth]{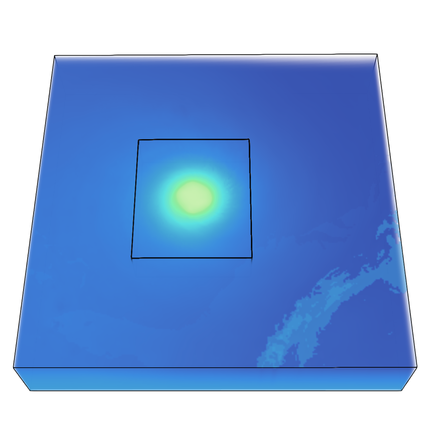}
    \caption{ROI for Isabel~data.}
    \label{fig:isabel_VOI}
\end{subfigure}
~~~
\begin{subfigure}[H]{0.24\linewidth}
     \centering
    \includegraphics[width=\linewidth]{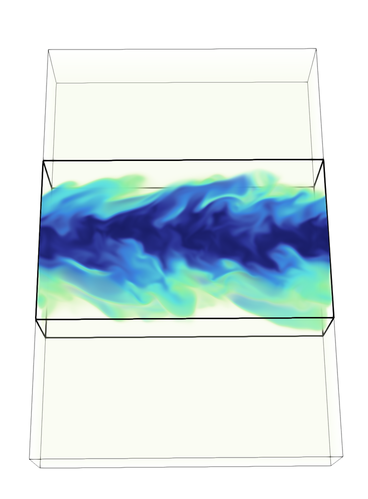}
    \caption{ROI for Combustion data.}
    \label{fig:combustion_VOI}
\end{subfigure}
~~~
\begin{subfigure}[H]{0.31\linewidth}
     \centering
    \includegraphics[width=\linewidth]{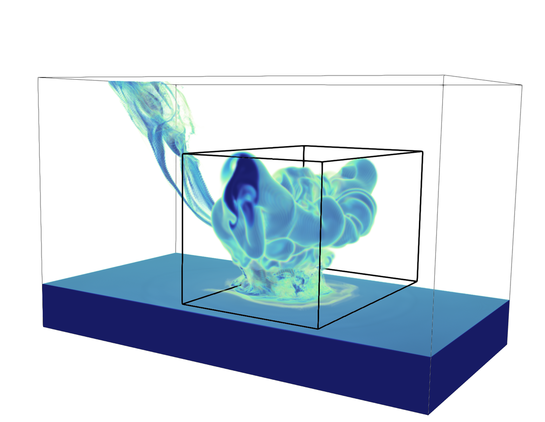}
    \caption{ROI for Asteroid~data.}
    \label{fig:asteroid_VOI}
\end{subfigure}
\caption{Regions of interest (ROI) of different data sets used for analysis. Figure~\ref{fig:correlation_VOI}a shows the ROI in Isabel data set, where the hurricane eye feature is selected. Figure~\ref{fig:correlation_VOI}b shows the ROI for Combustion data set, where the turbulent flame region is highlighted. Finally, in~Figure~\ref{fig:correlation_VOI}c the ROI for asteroid data set is shown. The~ROI selected in this example indicates the region where the asteroid has impacted the ocean surface and the splash of the water is ejected to the~environment.}
\label{fig:correlation_VOI}
\end{figure}
\unskip

\begin{table}[H]
\centering
\small
\caption{Evaluation of multivariate correlation for feature~regions.}
\label{corr_study}
\scalebox{0.9}{
\begin{tabular}{ccccccc}
\toprule
\multirow{2}{*}{} & \multicolumn{2}{c}{\textbf{Raw Data Correlation}} & \multicolumn{2}{c}{\textbf{PMI-Based Sampling}} & \multicolumn{2}{c}{\textbf{Random Sampling}} \\ \cmidrule{2-7} 
 & \begin{tabular}[c]{@{}c@{}}\textbf{Pearson's} \\ \textbf{Correlation}\end{tabular} & \begin{tabular}[c]{@{}c@{}}\textbf{Distance}\\ \textbf{Correlation}\end{tabular} & \begin{tabular}[c]{@{}c@{}}\textbf{Pearson's} \\ \textbf{Correlation}\end{tabular} & \begin{tabular}[c]{@{}c@{}}\textbf{Distance} \\ \textbf{Correlation}\end{tabular} & \begin{tabular}[c]{@{}c@{}}\textbf{Pearson's} \\ \textbf{Correlation}\end{tabular} & \begin{tabular}[c]{@{}c@{}}\textbf{Distance} \\ \textbf{Correlation}\end{tabular} \\ \midrule
\begin{tabular}[c]{@{}c@{}}Isabel Data\\ (Pressure and QVapor)\end{tabular} & $-$0.19803 & 0.3200 & $-$0.19805 & 0.3205 & $-$0.1966 & 0.3213 \\ \midrule
\begin{tabular}[c]{@{}c@{}}Combustion Data\\ (mixfrac and Y\_OH)\end{tabular} & 0.01088 & 0.4012 & 0.01624 & 0.4054 & 0.02123 & 0.4071 \\ \midrule
\begin{tabular}[c]{@{}c@{}}Asteroid Data\\ (tev and v02)\end{tabular} & 0.2116 & 0.2938 & 0.2273 & 0.2994 & 0.2382 & 0.31451 \\ \midrule
\end{tabular}}
\end{table}
\unskip

\begin{table}[H]
\centering
\small
\caption{Evaluation of multivariate correlation for full~data.}
\label{corr_study_full}
\scalebox{0.95}{
\begin{tabular}{ccccccc}
\toprule
\multirow{2}{*}{} & \multicolumn{2}{c}{\textbf{Raw Data}} & \multicolumn{2}{c}{\textbf{PMI-Based Sampling}} & \multicolumn{2}{c}{\textbf{Random Sampling}} \\ \cmidrule{2-7} 
 & \begin{tabular}[c]{@{}c@{}}\textbf{Pearson's} \\ \textbf{Correlation}\end{tabular} & \begin{tabular}[c]{@{}c@{}}\textbf{Distance}\\ \textbf{Correlation}\end{tabular} & \begin{tabular}[c]{@{}c@{}}\textbf{Pearson's} \\ \textbf{Correlation}\end{tabular} & \begin{tabular}[c]{@{}c@{}}\textbf{Distance} \\ \textbf{Correlation}\end{tabular} & \begin{tabular}[c]{@{}c@{}}\textbf{Pearson's} \\ \textbf{Correlation}\end{tabular} & \begin{tabular}[c]{@{}c@{}}\textbf{Distance} \\ \textbf{Correlation}\end{tabular} \\ \midrule
\begin{tabular}[c]{@{}c@{}}Isabel Data\\ (Pressure and QVapor)\end{tabular} & 0.3725 & 0.5470 & 0.3735 & 0.5530 & 0.3686 & 0.5480 \\ \midrule
\begin{tabular}[c]{@{}c@{}}Combustion Data\\ (mixfrac and Y\_OH)\end{tabular} & 0.3462 & 0.5113 & 0.3588 & 0.5248 & 0.3663 & 0.5321 \\ \midrule
\begin{tabular}[c]{@{}c@{}}Asteroid Data\\ (tev and v02)\end{tabular} & $-$0.028 & 0.3622 & $-$0.0209 & 0.1795 & $-$0.0259 & 0.1797 \\ \midrule
\end{tabular}}
\end{table}
}

\section{Discussion, Limitations, and~Future~Works}
\bmark{
In this work, we propose a new multivariate statistical association driven data sub-sampling algorithm that aims at preserving the statistically associated regions in the multivariate data sets with higher accuracy in the form of a reduced sub-sampled data set. To~achieve this, the~proposed technique first uses pointwise information theoretic measures to compute the strength of statistical association for each data point in the spatial domain and then uses a multivariate distribution-based approach to sample data points according to the strength of their multivariate association. Therefore, such a sampling technique produces a sub-sampled data set where stronger statistically associated regions are sampled densely compared to the regions that demonstrate weaker association among selected variables. We show several visualization and analysis tasks such as (a) multivariate feature query, (b) reconstruction based feature visualization and (c) recovery of multivariate correlation and dependence for important regions of interest in the data sets where the proposed sampling technique can be preferred over the traditional random sampling algorithms. This is primarily due to the fact that the random sampling technique samples the whole data set randomly and the importance of all the data points are equal. Hence, it does not sample any region~densely. 

However, when experts want to preserve the overall data distribution and statistical quantities like mean, standard deviation, random sampling technique can be preferred over the proposed method. Furthermore, as~we have observed from Table~\ref{corr_study_full}, when we are considering the full data, random sampling based technique is able to perform almost similarly to the proposed method in recovering the correlations. Therefore, it can be concluded that both the proposed sampling technique and the random sampling technique have their own advantages and limitations. The~proposed method does not recommend replacing random sampling completely, rather, they can be used in a complementary fashion depending on the demands of the application tasks. Hence, it can be concluded that the proposed technique is able to preserve the features (identified as statistically associated regions) in the scientific data sets more accurately compared to the random sampling methods, however, for~preserving the overall statistical data properties and distributions, random sampling can still be~used.

Another potential limitation of the proposed method in its current state is the use of multivariate histograms whose dimension increases with the number of variables used during sampling. When~the number of variables is large, creating a high-dimensional histogram can be challenging and also memory consuming. Therefore, to~overcome this issue, we plan to first cluster the variables using their shared information content into smaller sub-groups as was proposed in~\cite{vis13_biswas} and then apply the proposed sampling algorithm. Furthermore, we wish to compare the performance and efficacy of the proposed method with more sophisticated data sampling techniques such as stratified random sampling, importance driven sampling algorithms, etc.
}

In addition to the above goals, another future task is to apply the proposed sampling method to the distributed parallel environment such that the sampling algorithm can be run in~situ while the simulation is running and the generated data resides in the supercomputer memory. This will lead to an effective in~situ statistical sampling-based data summarization technique and the reduced sub-sampled data can be explored timely in the post-hoc analysis phase. In~order to perform the sampling in a distributed environment, we need to compute the global joint data distributions and we plan to leverage the optimized multivariate histogram computation technique provided in~\cite{kewei_multivar_histogram} for estimating compact multivariate distributions.  We also plan to apply our sampling algorithm to other scientific applications and obtain user feedback to further improve the quality of the sampling~technique.  
\section{Conclusions}

In summary, this paper presents a new multivariate association driven data sampling technique for summarization of large-scale scientific data sets. We use pointwise information theoretic measures to assign importance to data points based on their statistical association and then perform data sub-sampling according to such multivariate importance criterion. The~proposed sampling algorithm is applied to several scientific data sets to conduct multivariate feature analysis. Through comprehensive quantitative and qualitative query-driven analysis and reconstruction-based visualization, the~usefulness of the proposed sampling algorithm is justified.

\vspace{6pt}
\authorcontributions{ {Soumya Dutta was responsible for conceiving the idea, designing all the experiments, performing the studies, and writing
the paper. Ayan Biswas helped in conceiving the idea and in the editing of the manuscript. James Ahrens supervised the work and helped in guiding the research overall.} 
}

\funding{This research was funded by U.S. Department of Energy and Alpine Exascle Computing~Project.} 

\acknowledgments{The authors wish to thank the anonymous reviewers for their insightful and detailed comments. Also, the authors thank Francesca Samsel for her help with the colormaps. This research was supported by the Exascale Computing Project (ECP), Project Number: 17-SC-20-SC, a~collaborative effort of two DOE organizations - the Office of Science and the National Nuclear Security Administration. This work is published under LA-UR-19-24243 v2.}

\conflictsofinterest{The authors declare no conflict of interest for this~work.} 




\reftitle{References}



\end{document}